\documentclass[preprintnumbers, prd, twocolumn, floatfix, preprintnumbers, letterpaper, amsmath, amssymb, superscriptaddress]{revtex4}
\usepackage{amssymb, amsmath, bm, dcolumn, epsf, graphicx, latexsym, mathbbol, slashed, simplewick}
\usepackage{graphicx}
\usepackage{dcolumn}
\usepackage{bm}
\usepackage{amssymb}
\usepackage{color}
\usepackage{multirow}
\usepackage{makecell}
\usepackage[colorlinks,linkcolor=blue,citecolor=blue,urlcolor=blue]{hyperref}

\newcommand{\tc}{\textcolor{red}}

 \def\tskip{\setlength{\tskip}{5pt}}
\def\colwidth{\setlength{\colwidth}{3.5in}}

\newcommand{\lsim}{\mathrel{\hbox{\rlap{\lower.55ex\hbox{$\sim$}} \kern-.3em \raise.4ex \hbox{$<$}}}}
\newcommand{\gsim}{\mathrel{\hbox{\rlap{\lower.55ex\hbox{$\sim$}} \kern-.3em \raise.4ex \hbox{$>$}}}}
\newcommand{\beq}{\begin{equation}}
\newcommand{\eeq}{\end{equation}}
\newcommand{\be}{\begin{equation}}
\newcommand{\ee}{\end{equation}}
\newcommand{\bes}{\begin{equation*}}
\newcommand{\ees}{\end{equation*}}
\newcommand{\beqa}{\begin{eqnarray}}
\newcommand{\eeqa}{\end{eqnarray}}
\newcommand{\bea}{\begin{eqnarray}}
\newcommand{\ena}{\end{eqnarray}}

\begin{document}

\title{Post-Newtonian parameters and cosmological constant of screened modified gravity}

\author{Xing Zhang}
\email[]{starzhx@mail.ustc.edu.cn}
\author{Wen Zhao}
\email[]{wzhao7@ustc.edu.cn}
\author{He Huang}
\author{Yifu Cai}
\affiliation{CAS Key Laboratory for Researches in Galaxies and Cosmology, Department of Astronomy, University of Science and Technology of China, Chinese Academy of Sciences, Hefei, Anhui 230026, China}

\date{\today}


\begin{abstract}
Screened modified gravity (SMG) is a kind of scalar-tensor theories with screening mechanisms, which can generate screening effect to suppress the fifth force in high density environments and pass the solar system tests. Meanwhile, the potential of scalar field in the theories can drive the acceleration of the late universe. In this paper, we calculate the parameterized post-Newtonian (PPN) parameters $\gamma$ and $\beta$, the effective gravitational constant $G_{\rm eff}$ and the effective cosmological constant $\Lambda$ for SMG with a general potential $V$ and coupling function $A$. The dependence of these parameters on the model parameters of SMG and/or the physical properties of the source
object are clearly presented. As an application of these results, we focus on three specific theories of SMG (chameleon, symmetron and dilaton models). Using the formulae to calculate their PPN parameters and cosmological constant, we derive the constraints on the model parameters by combining the observations on solar system and cosmological scales.


\end{abstract}


\maketitle

\section{Introduction \label{section1}}

The current cosmic acceleration \cite{Supernova1998} can be elucidated within General Relativity (GR) by introducing the dark energy \cite{dark energy}.  Some prominent candidates for dark energy are the cosmological constant \cite{cc}, a dynamically evolving scalar field quintessence \cite{Quintessence}, a phantom field \cite{phantom}, and a quintom field \cite{quintom}, etc. Alternatively, the accelerated expansion of the universe can also be explained through modified gravity (MG) theories \cite{modified gravity}. On large scales, we do not have very strict experiments to verify GR, then infrared (IR) modification of gravity is the direction that is supposed to be worth a try \cite{Boisseau-2000PRD}. Weinberg's theorem states that any Lorentz invariant spin-2 field theory must reduce to GR at low-energy limit \cite{weinberg}, and thus any MG theories must involve extra degree(s) of freedom. The scalar degree of freedom universally exists in the fundamental physics (such as compactified extra dimensions \cite{Cho-1975-PRD}, string theory and brane world \cite{string}). Since Higgs boson in the Standard Model of particles was found \cite{higgs}, we know that scalar particles really exist in nature. Moreover, scalar fields are also widely used in cosmology. Quintessence scalar field can replace the cosmological constant and drive cosmic acceleration at late times \cite{Quintessence}. The inflation is a short period of rapid expansion in the very early universe, which could also be caused by a scalar field \cite{inflation.guth,inflation-reheating}. These scalar fields may couple to matter fields, which slightly violates GR and could be detected as the continuous improvements of experimental accuracy.

Most MG theories involve scalar field, and the simplest one is the so-called scalar-tensor gravity \cite{Scalar-Tensor,scalar-tensor Solar System,Scalar-tensor cosmologies,f(R),Brans-Dicke-PPN}. The fundamental building blocks of scalar-tensor theories are tensor gravitational field and scalar field. Moreover, scalar-tensor theories can be justified by the low-energy limit of string theory or supergravity \cite{String,Runaway Dilaton}. Scalar-tensor theories are usually expressed either in the Jordan frame or in the Einstein frame, which are related to each other by a conformal rescaling \cite{conformal frame}. In the Einstein frame, a key ingredient of scalar-tensor theories is the conformal coupling of light scalar field with matter fields, which usually implies the existence of a new long-range fifth force. However, at present, fifth forces have not been detected in either solar system or laboratory experiments, which means that the strength of fifth force should be much weaker than that of gravitational force \cite{Torsion balance experiments,fifth force}. Therefore, we need the screening mechanisms, which can suppress fifth force and allow MG theories to evade the tight gravitational tests in the solar system and the laboratory.

Examples of such screened models abound. The chameleon mechanism \cite{chameleon cosmology,chameleon fields,chameleon1,modified chamelons,chameleon-galaxy} operates a thin-shell shielding scalar field, which acquires a large mass in dense environments and suppresses its ability to mediate a fifth force. The symmetron mechanism \cite{symmetron-Khoury,symmetron cosmolgy,symmetron-Brax,symmetron-STRUCTURE,symmetron-galaxy,symmetron-modified} relies on the scalar field with the $\mathbb Z_2$ symmetry breaking potential. In high density regions the $\mathbb Z_2$ symmetry is unbroken and the fifth force is absent, whereas in low density regions the $\mathbb Z_2$ symmetry is spontaneously broken and the fifth force is present. The dilaton mechanism \cite{String,Dilaton and modified gravity,dilaton-Nonlinear structure formation,dilaton-Morris} is similar to the symmetron. The coupling between dilaton and matter is negligible in dense regions, while in low density regions the dilaton mediates a gravitational-strength fifth force. These screening mechanisms can be described by the same formalism \cite{Unified-SMG}, which is defined by a potential $V(\phi)$ and a coupling function $A(\phi)$ in scalar-tensor theory in the Einstein frame. Such scalar-tensor gravity with screening mechanism is often called screened modified gravity (SMG) \cite{Unified-SMG,SMG,SMG-cluster}. A basic requirement of SMG is that the effective potential must have a minimum \cite{Unified-SMG}, which can naturally be understood as a stable vacuum. This requirement can roughly constrain the shapes of two dynamical functions $V(\phi)$ and $A(\phi)$.

In this paper, we will focus on a generic SMG with arbitrary potential $V(\phi)$ and coupling function $A(\phi)$, and calculate the parameterized post-Newtonian (PPN) parameters $\gamma$ and $\beta$ in the case of a static spherically symmetric source. Moreover, SMG contains a scalar degree of freedom, whose potential can naturally provide the vacuum energy to drive the cosmic acceleration at late times. These two analyses allow us to investigate the theoretical framework on solar system and cosmological scales to derive the combined constraints on model parameters.


In the literature \cite{ppn-st-gen-potential,Testing-st-ppn}, the PPN parameters of a generic scalar-tensor theory were calculated under the assumption of point source surrounded by vacuum. This assumption is generally not appropriate to solve the massive scalar field, since the exterior scalar field of an extended source behaves quite different from that of a point source and screening mechanisms can show up due to non-linear effects of scalar field \cite{Testing-st-ppn}. So, these results are not applicable to the SMG, whose scalar field is always massive and can be screened in dense bodies.

In this paper, we solve the massive scalar field in the Einstein frame in the case of an extended source surrounded by a homogeneous background. Making use of this scalar solution and the PPN formalism \cite{Will:1993ns,WillReview}, in the Einstein frame we solve the massless metric field in the weak field limit around the flat Minkowski background and the vacuum expectation value (VEV) of the scalar field (scalar background). Then, we transform them to the Jordan frame and calculate the PPN parameters $\gamma$, $\beta$ and the effective gravitational constant $G_{\rm eff}$. It turns out that these parameters ($\gamma$, $\beta$, $G_{\rm eff}$) depend not only on the distance $r$ between the source object and the test mass, but also on the screened parameter $\epsilon$. 

Moreover, SMG contains a scalar degree of freedom, and the bare potential VEV of the scalar field can play the role of dark energy to accelerate the expansion of the universe. Further analysis shows that a generic SMG can converge back to GR with a cosmological constant in the limiting case $\epsilon\rightarrow 0$. In particular, we focus on three specific theories of SMG (chameleon, symmetron and dilaton models), and use our formulae to calculate their PPN parameters and cosmological constant, respectively. We find that our expressions of the PPN parameters for these three models can reduce to previous results derived by other authors in the appropriate cases. Finally, we combine solar system and cosmological constraints on these three models, respectively.

This paper is organized as follows. In Sec. \ref{section2}, we display the action and field equations for a generic SMG, and solve the scalar field equation in the Einstein frame. In Sec. \ref{section3}, we derive the post-Newtonian metric field equations in the Einstein frame. In Sec. \ref{section4}, we solve the post-Newtonian metric field equations, and calculate the PPN parameters and cosmological constant for a generic SMG. In Sec. \ref{section5}, we discuss chameleon, symmetron and dilaton models, respectively, and constrain them by the current observations. Finally, we conclude our results in Sec. \ref{section6}.

Throughout this paper, the metric convention is chosen as $(-,+,+,+)$, and Greek indices ($\mu,\nu,\cdots$) run over $0,1,2,3$. We set the units to $c=\hbar=1$, and therefore the reduced Planck mass is $M_\text{Pl} = \sqrt{1/8 \pi G}$, where $G$ is the Newtonian gravitational constant.

\section{Action functional and field equations\label{section2}}
A general scalar-tensor theory with two arbitrary functions is given by the following action in the Einstein frame \cite{Scalar-Tensor,scalar-tensor Solar System,Scalar-tensor cosmologies,f(R),Brans-Dicke-PPN}:
\begin{eqnarray} \label{action}
S_\text E&\!\!=\!\!\!&\int\!\! d^4x\sqrt{-g_\text E}\!\left[\frac{M_\text{Pl} ^2}{2}R_\text E\!-\!\frac12(\nabla_\text E\phi)^2\!-\!V(\phi)\right] \nonumber\\
&&+S_m\!\left[A^2\!(\phi) g^\text E_{\mu\nu},\,\psi_m^{(i)}\right],
\end{eqnarray}
where $g_\text {E}$ is the determinant of the Einstein frame metric $g^\text {E}_{\mu\nu}$, $R_\text {E}$ is the Ricci scalar, $\psi_m^{(i)}$ are various matter fields labelled by $i$, $V(\phi)$ is a bare potential characterizing the scalar self-interaction, and $A(\phi)$ is a conformal coupling function. In the Einstein frame, the scalar field $\phi$ interacts directly with matter fields $\psi_m^{(i)}$ through the conformal coupling function $A(\phi)$. In the Jordan frame, the matter fields $\psi_m^{(i)}$ couple to the Jordan frame metric $g^\text {J}_{\mu\nu}$ through a conformal rescaling of the Einstein frame metric $g^\text {E}_{\mu\nu}$ as \cite{conformal frame}
\be\label{conformal transformation}
g^\text {J}_{\mu\nu}=A^2(\phi)g^\text {E}_{\mu\nu},
\ee
where the coupling function $A(\phi)$ is usually different for different matter fields $\psi_m^{(i)}$. For simplicity, from now on, we assume that all matter fields couple in the same way to the scalar field with a universal coupling function $A(\phi)$.

The variation of the action \eqref{action} with respect to the metric field and the scalar field yields the metric field equation of motion (EOM) and the scalar field EOM:
%
\begin{align}
\label{tensor field eq.}
R^\text E_{\mu\nu}
&= 8 \pi G \Big[ S^\text E_{\mu\nu}+\partial_\mu\phi\partial_\nu \phi+V(\phi)g^\text E_{\mu\nu}\Big]
\\
\label{scalar field eq.1}
\Box\phi
&=\frac{{\rm d}V(\phi)}{{\rm d}\phi}-T^\text E
\frac{{\rm d}A(\phi)}{A(\phi){\rm d}\phi}
\end{align}
with
\be\label{Smunu}
S^\text E_{\mu\nu}\equiv T^\text E_{\mu\nu}- \frac{1}{2} g^{\text E}_{\mu\nu}T^\text E,
\ee
where $T^{\text E}_{\mu\nu} \equiv(-2/\sqrt{-g_{\text E}})\delta S_m/\delta g_\text E^{\mu\nu}$ is the energy-momentum tensor of matter in the Einstein frame, $T^\text E$ is the trace of the energy-momentum tensor $T_\text E^{\mu\nu}$, and $\Box\equiv g^{\mu\nu}_{\text E}\nabla_\mu\nabla_\nu $. The scalar field EOM \eqref{scalar field eq.1} can be rewritten as follows (Klein-Gordon equation),
\be\label{scalar field eq.2}
\Box \phi = \frac {{\rm d}V_{\rm eff}}{{\rm d}\phi}
\ee
with the effective potential
\be\label{Veff}
V_{\text {eff}}(\phi) \equiv V(\phi) +\rho \big[A(\phi)-1\big],
\ee
where the matter is assumed to be non-relativistic. $\rho$ is defined as the conserved energy density in the Einstein frame, i.e., $\rho$ is independent of $\phi$. The density $\rho$ is related to the Einstein frame and Jordan frame matter densities by \cite{Beyond Cos-Sta-Mod}
\be\label{phi-J-E}
\rho =\frac{\rho_{\text E}}{A}= A^3 \rho_{\text J}.
\ee

The scalar field is governed by the effective potential $V_{\text {eff}}(\phi)$, and the shape of the effective potential determines the behavior of the scalar field. For a general scalar-tensor theory with two arbitrary functions $V(\phi)$ and $A(\phi)$, the shape of the effective potential $V_{\text {eff}}(\phi)$ is usually arbitrary, and this scalar field generally does not have screening properties. For suitably chosen functions $V(\phi)$ and $A(\phi)$, the effective potential $V_{\rm eff}(\phi)$ can have a minimum, i.e., the scalar field has a physical vacuum. Around this minimum (physical vacuum), the scalar field acquires an effective mass which increases as the ambient density increases, and the scalar field can be screened in high density environments. This kind of scalar-tensor gravity with screening mechanism is often called the screened modified gravity (SMG) \cite{Unified-SMG,SMG,SMG-cluster}, which can generate screening effect to suppress the fifth force in high density environments and pass the solar system tests. There are many SMG models in the market, including the chameleon, symmetron and dilaton models \cite{Unified-SMG}, in which the functions $V(\phi)$ and $A(\phi)$ are chosen as the specific forms.


The following two conditions \eqref{phi min} and \eqref{m eff} guarantee that the effective potential $V_{\rm eff}(\phi)$ has a minimum. Differentiation of the effective potential with respect to $\phi$ is zero at $\phi=\phi_{\rm min}(\rho)$, i.e.
\begin{subequations}
\be\label{phi min}
\frac{\rm d V_{\rm eff}}{\rm d\phi}\big|_{\phi_{\rm min}}=0,
\ee
and the value of $\phi_{\rm min}(\rho)$ decreases as the ambient density increases.
The effective mass $m_{\rm eff}(\rho)$ of the scalar field at the minimum is defined as,
\be\label{m eff}
m^2_{\rm eff}\equiv \frac{\rm d^2 V_{\rm eff}}{\rm d\phi^2}\big|_{\phi_{\rm min}},
\ee
\end{subequations}
which should be a positive and monotonically increasing function of the ambient density.

Let us consider a static spherically symmetric and constant density source object, which is embedded in a homogeneous background. Then, the scalar field EOM \eqref{scalar field eq.2} simplifies to
%
\be
\label{scalar field eq.3}
\frac{{\rm d}^2\phi}{{\rm d}r^2}+\frac 2r\frac{{\rm d}\phi}{{\rm d}r}=m^2_{\rm m}(\rho)\big[\phi-\phi_{\rm m}(\rho)\big]
\ee
with
\begin{equation}
\rho(r) = \left\{
\begin{matrix}
\rho_0\qquad~{\rm for}~~~ r<R \cr
\rho_{\scriptscriptstyle\!\infty}\qquad~ {\rm for}~~~ r>R
\end{matrix}
\right.,
\end{equation}
where $R$ is the radius of the source object, $\rho_0$ is the density of the source object, and $\rho_{\scriptscriptstyle\!\infty}$ is the background matter density. For the solar system, in general, $\rho_{\scriptscriptstyle\!\infty}$ is the cosmological matter density or galactic matter density \cite{Combined-cc-ss,Unified-SMG}, which corresponds to the cosmological background or galactic background, respectively.

Eq.~\eqref{scalar field eq.3} is a second order differential equation, and the boundary conditions are required as follows \cite{chameleon cosmology},
\begin{align}
\begin{split}
& \frac{\rm d\phi}{\text dr} = 0 ~\,\qquad {\rm at}~~~r=0~ \\
& \phi\rightarrow \phi_{\scriptscriptstyle\!\infty} ~\qquad {\rm as}~~~r\!\rightarrow\! \infty\,,
\end{split}
\end{align}
where $\phi_{\scriptscriptstyle\!\infty}$ is the scalar field VEV (scalar background), depending on the background matter density $\rho_{\scriptscriptstyle\!\infty}$. The first condition guarantees that the scalar field is non-singular at the origin \cite{chameleon cosmology}, and the second one implies that the scalar field asymptotically converges to the scalar background. Moreover, $\phi$ and ${\rm d}\phi/{\rm d}r$ are of course continuous at the surface of the source object. By solving Eq. \eqref{scalar field eq.3} directly, we get the exact solution

\begin{subequations}
\begin{align}
\phi(r<R)&=\phi_0+\frac Ar\sinh(m_0r)
\\
\label{exterior scalar field}
\phi(r>R)&=\phi_{\scriptscriptstyle\!\infty}+\frac Bre^{-m_{\scriptscriptstyle\!\infty} r}
\end{align}
\end{subequations}
with
\begin{subequations}
\begin{align}
A&=\frac{(\phi_{\scriptscriptstyle\!\infty}-\phi_0)(1+m_{\scriptscriptstyle\!\infty} R)}{m_0\cosh(m_0R)+m_{\scriptscriptstyle\!\infty}\sinh(m_0R)}
\\
B&=-e^{m_{\scriptscriptstyle\!\infty} R}(\phi_{\scriptscriptstyle\!\infty}-\phi_0)\frac{m_0R-\tanh(m_0R)}{m_0+m_{\scriptscriptstyle\!\infty}\tanh(m_0R)},
\end{align}
\end{subequations}
where $\phi_0$ and $\phi_{\scriptscriptstyle\!\infty}$ are respectively the positions of the minimum of $V_{\rm eff}$ inside and far outside the source object, $m_0$ and $m_{\scriptscriptstyle\!\infty}$ are respectively the effective masses of the scalar field at $\phi_0$ and $\phi_{\scriptscriptstyle\!\infty}$. All these quantities can be obtained by two given functions $V(\phi)$ and $A(\phi)$.

The scalar field is screened on solar system scales (high density), which requires that the typical scale of the solar system $R$ is much larger than the fifth force range $m^{-1}_0$. In addition, the scalar field works on cosmological scales (low density), which requires that $m^{-1}_{\scriptscriptstyle\!\infty}$ is close to the Hubble scale. So, the conditions $m_0R\gg1$ and $m_{\scriptscriptstyle\!\infty} R\ll1$ can always be satisfied on solar system scales. In this paper, we only consider the exterior solution of the scalar field. Using these two relations, the exterior scalar field \eqref{exterior scalar field}  is reduced to
%
\begin{align}\label{exterior scalar solution}
\begin{split}
\phi(r)
&=\phi_{\scriptscriptstyle\!\infty}-\epsilon M_\text{Pl}\frac{GM_{\text E}}re^{-m_{\scriptscriptstyle\!\infty} r}
\end{split}
\end{align}
with
\be\label{epsilon}
\epsilon\equiv\frac{\phi_{\scriptscriptstyle\!\infty}-\phi_0}{M_\text{Pl}\Phi_{\text E}},
\ee
where $M_\text{E}$ is the mass of the source object in the Einstein frame, $\Phi_{\text E}\equiv GM_\text{E}/R$ is the Newtonian potential at the surface of the source object in the Einstein frame, and the parameter $\epsilon$ depends on background matter density $\rho_{\scriptscriptstyle\!\infty}$ and the physical properties (density  $\rho_0$ and radius $R$) of the source object. Obviously, the screening effect is very strong for $\epsilon\ll1$, and quite weak for $\epsilon\gtrsim1$, so $\epsilon$ is always called the screened parameter or the thin-shell parameter in the literature \cite{chameleon cosmology}.

This completes the solution of the scalar field EOM in the Einstein frame. In the next section, we will use the scalar field solution to derive the post-Newtonian metric field equations in the Einstein frame.

\section{Post-Newtonian approximation in the Einstein frame\label{section3}}
In order to solve the metric field EOM \eqref{tensor field eq.}, we make use of the PPN formalism introduced in \cite{Will:1993ns, WillReview}. In this formalism, the gravitational field of the source is weak $GM/r\ll1$, and the typical velocity $\vec v$ of the source matter is small $v^2\sim GM/r\ll1$. Thus, we can use the perturbative expansion method to solve the field equations, and all dynamical quantities can be expanded to $\mathcal{O}(n) \propto {v}^{2n}$ (Note that, other authors use the convention $\mathcal{O}(n) \propto {v}^{n}$).

In this section, we consider a static spherically symmetric source, and assume that the source object is constituted by a perfect fluid which obeys the post-Newtonian hydrodynamics. We start from this assumption, and expand the metric field EOM to  $\mathcal{O}(n) \propto {v}^{2n}$ in the weak field limit around the flat Minkowski background and the scalar background (scalar field VEV). The resulting equations can then be solved subsequently for each order of magnitude in the next section.

For the metric field $g_{\mu\nu}$ in the weak field, it can be expanded around the flat Minkowski background as follows,
\begin{align}
\begin{split}
g_{\mu\nu}
&=\eta_{\mu\nu}+h_{\mu\nu}
\\
&=\eta_{\mu\nu}+h^{(1)}_{\mu\nu}+h^{(2)}_{\mu\nu}+\mathcal{O}(3).
\end{split}
\end{align}
This metric can also be written in the spherically symmetric and isotropic coordinates $(t_{\text E}, r, \theta, \varphi )$ in the Einstein frame,
\begin{align}\label{EF metric}
\begin{split}
ds^2_\text E=&-\left[ 1 - h_{\text E00}^{(1)}(r)
 - h_{\text E00}^{(2)}(r)\right] dt^2_\text E
\\&+ \left[1+h_{\text Err}^{(1)}(r)\right]
\left( dr^2 + r^2 d\Omega^2 \right),
\end{split}
\end{align}
where $t_\text E$ and $r$ are the time and radial coordinates in the Einstein frame, respectively. Each term $h^{(n)}_{\text E\mu\nu}$ is of order $\mathcal{O}(n)$. The term $d\Omega^2$ is defined by $d\Omega^2\equiv dr^2+\sin^2\theta d\varphi^2$, and the flat Minkowski background is $\eta_{\mu\nu} = \text{diag}\big(-1,1,r^2,r^2 \sin^2\theta\big)$.

For the scalar field $\phi$, the exterior solution \eqref{exterior scalar solution} is a following expansion in the weak field limit around the scalar background,
\be
\phi(r)= \phi_{\scriptscriptstyle\!\infty} + \phi^{(1)}(r)
\ee
where $\phi^{(1)}$ is of order $\mathcal{O}(1)$, given by
\be
\label{scalar perturbation}
\phi^{(1)}(r)=-\epsilon M_\text{Pl}\frac{GM_{\text E}}re^{-m_{\scriptscriptstyle\!\infty} r},
\ee
%
$\phi_{\scriptscriptstyle\!\infty}\equiv\phi_{\scriptscriptstyle\rm \!V\!E\!V}$ is the scalar background (scalar field VEV), which depends on the background matter density $\rho_{\scriptscriptstyle\!\infty}$. Note that, the term $\phi^{(2)}$ naturally does not exist in our expression of the scalar field $\phi$ \eqref{exterior scalar solution}, which is different from the results derived in other method \cite{ppn-st-gen-potential,Testing-st-ppn}. Then, the bare potential $V(\phi)$ and the coupling function $A(\phi)$ can be expanded in Taylor's series around the scalar background,
\begin{subequations} \label{VA-expand}
\begin{align}
V(\phi) \label{V-expand}
&\!= \!V_{\scriptscriptstyle\rm \!V\!E\!V} \!+\! V_1 \left(\phi-\phi_{\scriptscriptstyle\!\infty}\right) \!+\! V_2 \left(\phi-\phi_{\scriptscriptstyle\!\infty}\right)^2\!+\!\mathcal{O}(3),
\\
A(\phi)\label{A-expand}
&\!=\! A_{\scriptscriptstyle\rm \!V\!E\!V}\!+\! A_1\! \left(\phi-\phi_{\scriptscriptstyle\!\infty}\right) \!+\! A_2 \!\left(\phi-\phi_{\scriptscriptstyle\!\infty}\right)^2\!\!+\!\mathcal{O}(3),
\end{align}
\end{subequations}
where $A_{\scriptscriptstyle\rm \!V\!E\!V}\equiv A(\phi_{\scriptscriptstyle\rm \!V\!E\!V})$ is the coupling function VEV, and $V_{\scriptscriptstyle\rm \!V\!E\!V}\equiv V(\phi_{\scriptscriptstyle\rm \!V\!E\!V})$ is the bare potential VEV, which acts as the effective cosmological constant to accelerate the expansion of the late universe.

The energy-momentum tensor is given by that of a perfect fluid \cite{Will:1993ns,WillReview}
\begin{align}
T^{\mu\nu} = \left( \rho + \rho \Pi + p \right) u^\mu u^\nu
+ p g^{\mu\nu},
\end{align}
and the tensor $S_{\mu\nu}^\text E$ \eqref{Smunu} is expanded in the form:
\begin{subequations}
\begin{align}
\begin{split}
S^\text E_{00} =&\frac{1}{2}\rho_\text E\big(1\!+\!\Pi_\text E\!+\!2v^2_\text E\!-\!h_{\text E00}^{(1)}\big)\!+\!\frac{3}{2}p_\text E\!+\! \mathcal{O}(3)
\end{split}
\\
S^\text E_{rr} =&\frac{1}{2}\rho_\text E
+ \mathcal{O}(2)
\\
S^\text E_{\theta\theta} =&\frac{1}{2}\rho_\text Er^2
+ \mathcal{O}(2)
\\
S^\text E_{\varphi\varphi} =&\frac{1}{2}\rho_\text Er^2\sin^2\theta
+ \mathcal{O}(2),
\end{align}
\end{subequations}
where $\rho$ is density of rest mass, $p$ is pressure, $\Pi$ is internal energy per unit rest mass, $u^\mu$ is four-velocity, and the index $\rm E$ indicates that a quantity is defined in the Einstein frame. For solar system tests, we typically have $p\ll\rho$, $\Pi\ll1$ and $v\ll1$. So, we neglect the effects of pressure, internal energy and velocity in the following discussions.

By using these relations, the right-hand sides of the metric field EOM \eqref{tensor field eq.} can be expanded to the required order in the form:
\begin{subequations}\label{R-right}
\begin{align}
\begin{split}
R^\text E_{00}
 \!=&8\pi \!G\!\bigg[\!\!-\!V_{\scriptscriptstyle\rm \!V\!E\!V}\!+\!\frac{\rho_\text E}2\!+\!\!V_{\scriptscriptstyle\rm \!V\!E\!V}h_{\text E00}^{(1)}\!\! -\!\!V_1\phi^{(1)}\!\!-\!\frac{\rho_\text E}2h_{\text E00}^{(1)}
\\&+\!V_{\scriptscriptstyle\rm \!V\!E\!V}h_{\text E00}^{(2)}\!\!+\!V_1\phi^{(1)}h_{\text E00}^{(1)}\!\!-\!V_2\big(\phi^{(1)}\big)^2\bigg]\!+\!\mathcal{O}(3)\label{R00-right}
\end{split}
\\
\begin{split}
R^\text E_{rr}
\!=&8\pi \!G\!\bigg[\!V_{\scriptscriptstyle\rm \!V\!E\!V}\!+\!\frac{\rho_\text E}2\!+\!V_{\scriptscriptstyle\rm \!V\!E\!V}h_{\text Err}^{(1)} \!+\!V_1\phi^{(1)} \!\bigg]\!+\!\mathcal{O}(2)\label{R11-right}
\end{split}
\\
\begin{split}
R^\text E_{\theta\theta}
\!=&8\pi \!G r^2\!\bigg[V_{\scriptscriptstyle\rm \!V\!E\!V}\!+\!\frac{\rho_\text E}2\!+\!V_{\scriptscriptstyle\rm \!V\!E\!V}h_{\text Err}^{(1)} \!+\!V_1\phi^{(1)}\! \bigg]\!\!+\!\mathcal{O}(2)\label{R22-right}
\end{split}
\\
R^\text E_{\varphi\varphi}
\!=&R^\text E_{\theta\theta} \sin^2\theta+ \mathcal{O}(2)\label{R33-right}.
\end{align}
\end{subequations}
The left-hand sides of the metric field EOM \eqref{tensor field eq.}, i.e., the components of the Ricci tensor, are expanded to the same order in the form:
\begin{subequations}\label{R-left}
\begin{align}
\begin{split}
R^\text E_{00}
\!=&\! -\! \frac{1}{2} \nabla_r^2 h_{\text E00}^{(1)}\!-\! \frac{1}{2}\! \bigg(\! \nabla_r^2 h_{\text E00}^{(2)}\!-\! h_{\text Err}^{(1)}  \nabla_r^2 h_{\text E00}^{(1)}
\\&
+\! \frac{1}{2}\!  ( \partial_r h_{\text E00}^{(1)})^2\!+\! \frac{1}{2}\! ( \partial_r h_{\text E00}^{(1)} )\! ( \partial_r h_{\text Err}^{(1)}\! )\!\bigg)\!+\! \mathcal{O}(3)\label{R00-left}
\end{split}
\\
R^\text E_{rr}
 \!=&\frac{1}{2} \partial_r^2 h_{\text E00}^{(1)}\!-\! \partial_r^2 h_{\text Err}^{(1)} \!-\! \frac{1}{r} \partial_r h_{\text Err}^{(1)}+ \mathcal{O}(2)\label{R11-left}
\\
R^\text E_{\theta\theta}
 \!=& \frac{1}{2} r^2 \!\left(\!\frac{1}{r} \partial_r h_{\text E00}^{(1)} \!-\! \partial_r^2 h_{\text Err}^{(1)}\!-\! \frac{3}{r} \partial_r h_{\text Err}^{(1)}\!\right)\!+\! \mathcal{O}(2)\label{R22-left}
\\
R^\text E_{\varphi\varphi}
\!=&R^\text E_{\theta\theta} \sin^2\theta+ \mathcal{O}(2)\label{R33-left},
\end{align}
\end{subequations}
where $\nabla_r^2\equiv\partial_r^2+2/r\partial_r$ is the flat space spherical coordinate Laplace operator. Obviously, Eq. (\ref{R33-right}, \ref{R33-left}) is equivalent to Eq. (\ref{R22-right}, \ref{R22-left}), so there are only three independent equations, which will be solved to derive the PPN parameters in the following sections.


\section{Static spherically symmetric solution\label{section4}}

\subsection{Metric in the Einstein frame}
Since the metric gravitational field is always massless in SMG, similar to the previous work \cite{ppn-st-gen-potential}, we solve the metric field equations in the case of a point source, i.e., $\rho_{\rm E}=M_{\rm E}\delta(r)$. In the following calculation, we neglect dark energy $V_{\scriptscriptstyle\rm \!V\!E\!V}$ \eqref{cc} and gravity $h^{(n)}_{\mu\nu}$ interaction terms $V_{\scriptscriptstyle\rm \!V\!E\!V}h^{(n)}_{\mu\nu}$, since the effect of this interaction is quite weak on solar system scales.

We consider the post-Newtonian metric field equation (\ref{R00-right}, \ref{R00-left}) up to first order, and obtain the equation
\begin{align}\label{h00 first order equation}
\begin{split}
\nabla^2_rh^{(1)}_{\text E00}=&8\pi G\Big(2V_{\scriptscriptstyle\rm \!V\!E\!V}-{\rho_\text E} +2V_1\phi^{(1)}\Big).
\end{split}
\end{align}
Using the scalar field \eqref{scalar perturbation}, the solution is given by
\begin{align}\label{hE001}
\begin{split}
h^{(1)}_{\text E00}\!(r)\!=&\frac{2G\!M_\text E}r\!\bigg(\!1\!-\!\frac{{V_1}}{M_{\rm \! Pl}m^2_{\scriptscriptstyle\!\infty}}\epsilon e^{\!-m_{\scriptscriptstyle\!\infty}\! r}\!\bigg)\!+\!\frac{8\pi \!GV_{\scriptscriptstyle\rm \!V\!E\!V}}3r^2.
\end{split}
\end{align}

For the spatial components, up to first order, the post-Newtonian metric field equations (\ref{R11-right}, \ref{R11-left}) and (\ref{R22-right}, \ref{R22-left}) follows that,
\begin{subequations}
\begin{align}
\frac{1}{2} \partial_r^2 h_{\rm E00}^{(1)}\!\!-\!\partial_r^2 h_{\text Err}^{(1)}\!\!-\!\frac{1}{r} \partial_r h_{\text Err}^{(1)}&\!\!=\!8\pi \!G\!\Big(\!V_{\scriptscriptstyle\rm \!V\!E\!V}\!+\!\frac{\rho_\text E}2\!+\!V_1\phi^{(1)} \!\Big)
\\
\frac{1}{r} \partial_r h_{\text E00}^{(1)} \!\!- \!\partial_r^2 h_{\text Err}^{(1)}\!\!-\! \frac{3}{r} \partial_r h_{\text Err}^{(1)}&\!\!=\!8\pi \!G\!\Big(\!2V_{\scriptscriptstyle\rm \!V\!E\!V}\!+\!\rho_\text E\!+\!2V_1\phi^{(1)}\! \Big).
\end{align}
\end{subequations}
Combining these two equations, and using Eq. \eqref{h00 first order equation}, we have
\begin{align}
\begin{split}
\nabla^2_rh^{(1)}_{\text Err}=&8\pi G\Big(-V_{\scriptscriptstyle\rm \!V\!E\!V}-{\rho_\text E} -V_1\phi^{(1)}\Big),
\end{split}
\end{align}
and the solution is also derived by applying the solution of scalar field in \eqref{scalar perturbation},
\begin{align}\label{hE111}
\begin{split}
h^{(1)}_{\text Err}\!(r)\!=&\frac{2G\!M_\text E}r\!\bigg(\!\!1\!+\!\frac{{V_1}}{2M_{\rm \! Pl}m^2_{\scriptscriptstyle\!\infty}}\epsilon e^{\!-m_{\scriptscriptstyle\!\infty}\! r}\!\!\bigg)\!-\!\frac{8\pi \!GV_{\scriptscriptstyle\rm \!V\!E\!V}}6r^2.
\end{split}
\end{align}

We now consider the post-Newtonian metric field equation (\ref{R00-right}, \ref{R00-left}). Up to second order, we obtain the equation
\begin{align}\label{hE00-2}
\begin{split}
&\nabla_r^2 h_{\text E00}^{(2)}+ \frac{1}{2}\partial_r h_{\text E00}^{(1)}\partial_r\big(h_{\text E00}^{(1)}+h_{\text Err}^{(1)}\big)
\\=&8\pi G\Big[2V_1\phi^{(1)}\big(h_{\text Err}^{(1)}-h_{\text E00}^{(1)}\big)+2V_2\big(\phi^{(1)}\big)^2\Big],
\end{split}
\end{align}
where we have neglected the terms $\rho_{\rm E} h^{(1)}_{\mu\nu}$ and $\rho_{\rm E}\phi^{(1)}$, which correspond to the gravitational self-energies and do not affect the calculation of the PPN parameter $\beta$ \cite{ppn-st-gen-potential,WillReview,Will:1993ns}. Using the metric fields \eqref{hE001} \eqref{hE111} and the scalar field \eqref{scalar perturbation}, the solution of Eq. \eqref{hE00-2} is given by
\begin{align}\label{hE002}
\begin{split}
h_{\text E00}^{(2)}\!(r)\!=\!&-\!\frac{2G^2\!M^2_\text E}{r^2}\!
\bigg[
\!1\!-\!\frac{5 V_1}{4M_\text {Pl}m^2_{\scriptscriptstyle\!\infty}}\!\Big(\!1\!-\!m_{\scriptscriptstyle\!\infty} r\!\Big)\epsilon e^{-m_{\scriptscriptstyle\!\infty} r}
\\&\!-\!\frac{V_2}{2m^2_{\scriptscriptstyle\!\infty}}\!\Big(\!1\!-\!\frac{3V^2_1}{M^2_{\text {Pl}}m^2_{\scriptscriptstyle\!\infty} V_2}\!\Big)\epsilon^2m_{\scriptscriptstyle\!\infty} re^{-2m_{\scriptscriptstyle\!\infty} r}
\\&\!+\!\frac{V^2_1}{4M^2_\text {Pl}m^4_{\scriptscriptstyle\!\infty}}\!\Big(\!1\!-\!m_{\scriptscriptstyle\!\infty} r\!\Big)\epsilon^2e^{-2m_{\scriptscriptstyle\!\infty} r}
\\&\!+\!\frac{5 V_1}{4M_\text{Pl}m^2_{\scriptscriptstyle\!\infty}}\epsilon(m_{\scriptscriptstyle\!\infty} r)^2\text{Ei}(\!-m_{\scriptscriptstyle\!\infty} r)
\\&\!-\!\Big(\!\frac{V_2}{m^2_{\scriptscriptstyle\!\infty}}-\!\frac{5V^2_1}{2M^2_\text{Pl}m^4_{\scriptscriptstyle\!\infty}}\!\Big)\epsilon^2(m_{\scriptscriptstyle\!\infty} r)^2\text{Ei}(\!-2m_{\scriptscriptstyle\!\infty} r)
\!\bigg],
\end{split}
\end{align}
where the function ${\rm Ei}(-x)$ is defined by the exponential integral
\begin{align}\label{Ei}
\text{Ei}(-x)\equiv - \int_x^\infty da \frac{e^{-a}}{a}.
\end{align}
The quantity $m_{\scriptscriptstyle\!\infty}$ is the effective mass of the scalar field at $\rho=\rho_{\scriptscriptstyle\!\infty}$.
Using the relations \eqref{VA-expand} and \eqref{m eff}, this quantity can be written as
\be\label{mV2A2}
m^2_{\scriptscriptstyle\!\infty}=2(V_2+\rho_{\scriptscriptstyle\!\infty}A_2).
\ee

\subsection{Metric in the Jordan frame}
SMG theories are usually expressed either in the Einstein frame or in the Jordan frame, and these two frames are related by a conformal rescaling \cite{conformal frame}. The PPN parameters are defined in the Jordan frame \cite{Will:1993ns,WillReview,ppn-st-gen-potential}, so we should transform to the Jordan frame to get the expressions of parameters $\gamma$ and $\beta$.

In the weak field limit, the metric is written in the spherically symmetric and isotropic coordinates $(t_{\text J}, \chi, \theta, \varphi )$ as follows,
\begin{align}\label{JF metric}
\begin{split}
ds^2_\text J=&-\left[ 1 - h_{\text J00}^{(1)}(\chi)
 - h_{\text J00}^{(2)}(\chi)\right] dt^2_\text J
\\&+ \left[1+h_{\text J\chi\chi}^{(1)}(\chi)\right]
\left( d\chi^2 + \chi^2 d\Omega^2 \right),
\end{split}
\end{align}
where $d\Omega^2\equiv dr^2+\sin^2\theta d\varphi^2$, $t_\text J$ and $\chi$ are the time and radial coordinates in the Jordan frame respectively, which relate to the corresponding quantities in the Einstein frame through the relations \eqref{t-chi-J}. This metric naturally satisfies the standard post-Newtonian gauge \cite{Will:1993ns}, and the PPN parameters $\gamma$ and $\beta$ are defined in the form \cite{Will:1993ns,WillReview}:
\begin{subequations}\label{JF metric component}
\begin{align}
\label{JF metric component00-1}
h_{\text{J}00}^{(1)}(\chi) &\equiv \frac{2 G_{\rm eff}(\chi) M_\text{J}}{\chi},
\\
\label{JF metric component11-1}
h_{\text{J}\chi\chi}^{(1)}(\chi) &\equiv  \gamma(\chi)\frac{2 G_{\rm eff}(\chi),
M_\text{J}}{\chi}
\\
\label{JF metric component00-2}
h_{\text{J}00}^{(2)}(\chi) &\equiv  -\beta(\chi) \frac{4 G_{\rm eff}^2(\chi)
M^2_\text{J}}{2\chi^2},
\end{align}
\end{subequations}
where $G_{\rm eff}$ is the effective gravitational `constant', and $M_\text J$ is the mass of the source object in the Jordan frame, which relates to the mass in the Einstein frame through the relation \eqref{MJE}. As mentioned above, in this paper, we neglect the effects of the pressure $p$, internal energy $\Pi$ and velocity $v$ of the source object, which may contribute additional PPN parameters \cite{Will:1993ns,WillReview,ppn-st-gen-potential}.


Using the relations \eqref{EF metric} and \eqref{A-expand}, the conformal rescaling \eqref{conformal transformation} turns into
\begin{align}\label{J-E-F}
\begin{split}
ds^2_\text{J}\! =& A^2(\phi) ds^2_\text{E}
\\
=& \!-\!\bigg[ 1 \!-\!h_{\text{E}00}^{(1)} \!+\! \frac{2A_1}{A_{\scriptscriptstyle\rm \!V\!E\!V}} \phi^{(1)}\!-\!h_{\text{E}00}^{(2)}
\!-\!\frac{2A_1}{A_{\scriptscriptstyle\rm \!V\!E\!V}} h_{\text{E}00}^{(1)} \phi^{(1)}
\\& + \left( \frac{2A_2}{A_{\scriptscriptstyle\rm \!V\!E\!V}} + \frac{A_1^2}{A_{\scriptscriptstyle\rm \!V\!E\!V}^2} \right)\big(\phi^{(1)}\big)^2 \bigg] {A_{\scriptscriptstyle\rm \!V\!E\!V}^2}dt_\text{E}^2
\\&+\!\left(1\! +\! h_{\text{E}rr}^{(1)} \!+\! \frac{2A_1}{A_{\scriptscriptstyle\rm \!V\!E\!V}} \phi^{(1)} \right)  {A_{\scriptscriptstyle\rm \!V\!E\!V}^2}\!\!\left({dr^2} +r^2 d\Omega^2\right).
\end{split}
\end{align}
Comparing this relation \eqref{J-E-F} with the Jordan frame metric in \eqref{JF metric}, we obtain the relations
\begin{subequations}\label{metric EF to JF}
\begin{align}
h_{\text{J}00}^{(1)}
\!=&h_{\text{E}00}^{(1)} - \frac{2A_1}{A_{\scriptscriptstyle\rm \!V\!E\!V}}  \phi^{(1)},
\\
h_{\text{J}\chi\chi}^{(1)}
\!=&h_{\text{E}rr}^{(1)} + \frac{2A_1}{A_{\scriptscriptstyle\rm \!V\!E\!V}} \phi^{(1)},
\\
\begin{split}
h_{\text{J}00}^{(2)}
\!=&h_{\text{E}00}^{(2)}\!+\! \frac{2A_1}{A_{\scriptscriptstyle\rm \!V\!E\!V}}h_{\text{E}00}^{(1)} \phi^{(1)}
\!\!-\! \left(\! \frac{2A_2}{A_{\scriptscriptstyle\rm \!V\!E\!V}} \!+\! \frac{A_1^2}{A_{\scriptscriptstyle\rm \!V\!E\!V}^2} \!\right)\!\!\big(\!\phi^{(1)}\!\big)^2,
\end{split}
\end{align}
with
\begin{align}
\begin{split}
t_\text{J} &= A_{\scriptscriptstyle\rm \!V\!E\!V}t_\text{E}
\\
\chi &= A_{\scriptscriptstyle\rm \!V\!E\!V}r.\label{t-chi-J}
\end{split}
\end{align}
\end{subequations}
Using the relations in \eqref{t-chi-J} and \eqref{phi-J-E}, the masses in these two frames are related by
\be\label{MJE}
M_{\rm J}=\frac{M_{\rm E}}{A_{\scriptscriptstyle\rm \!V\!E\!V}},
\ee
which follows the relation $M_{\rm J}\chi=M_{\rm E}r$.
%

Using the scalar field \eqref{scalar perturbation} and the metric fields \eqref{hE001} \eqref{hE111} \eqref{hE002}, from the relations \eqref{metric EF to JF}  we obtain the components of the Jordan frame metric:
\begin{widetext}
\begin{subequations}\label{hJ}
\begin{align}
h^{(1)}_{\text J00}(r)\label{hJ00-1}
=&\frac{2GM_\text E}r+\Big(\frac{A_1M_\text {Pl}}{A_{\scriptscriptstyle\rm \!V\!E\!V}}-\frac{V_1}{M_\text {Pl}m^2_{\scriptscriptstyle\!\infty}}\Big)\epsilon\frac{2GM_\text E}r e^{-m_{\scriptscriptstyle\!\infty} r}+\frac{8\pi GV_{\scriptscriptstyle\rm \!V\!E\!V}}3r^2,
\\
h^{(1)}_{\text J\chi\chi}(r)\label{hJ11-1}
=&\frac{2GM_\text E}r-\Big(\frac{A_1M_\text {Pl}}{A_{\scriptscriptstyle\rm \!V\!E\!V}}-\frac{V_1}{2M_\text {Pl}m^2_{\scriptscriptstyle\!\infty}}\Big)\epsilon\frac{2GM_\text E}r e^{-m_{\scriptscriptstyle\!\infty} r}-\frac{8\pi GV_{\scriptscriptstyle\rm \!V\!E\!V}}6r^2,
\\
\begin{split}
h_{\text J00}^{(2)}(r)\label{hJ00-2}
=&-\frac{2G^2M^2_\text E}{r^2}\bigg[1\!+\!\frac{2A_1M_\text {Pl}}{A_{\scriptscriptstyle\rm \!V\!E\!V}}\epsilon e^{-m_{\scriptscriptstyle\!\infty} r}\!-\!\frac{5 V_1}{4M_\text {Pl}m^2_{\scriptscriptstyle\!\infty}}\Big(1-m_{\scriptscriptstyle\!\infty} r\Big)\epsilon e^{-m_{\scriptscriptstyle\!\infty} r}
\!+\!M^2_\text{Pl}\!\Big(\frac{A^2_1}{2A^2_{\scriptscriptstyle\rm \!V\!E\!V}}\!+\!\frac{A_2}{A_{\scriptscriptstyle\rm \!V\!E\!V}}\!\Big)\epsilon^2e^{-2m_{\scriptscriptstyle\!\infty} r}
\\&-\frac{2V_1A_1}{m^2_{\scriptscriptstyle\!\infty}A_{\scriptscriptstyle\rm \!V\!E\!V}}\epsilon^2e^{-2m_{\scriptscriptstyle\!\infty} r}+\frac{V^2_1}{4M^2_\text {Pl}m^4_{\scriptscriptstyle\!\infty}}\Big(1-m_{\scriptscriptstyle\!\infty} r\Big)\epsilon^2e^{-2m_{\scriptscriptstyle\!\infty} r}-\frac{V_2}{2m^2_{\scriptscriptstyle\!\infty}}\Big(1-\frac{3V^2_1}{M^2_{\text {Pl}}m^2_{\scriptscriptstyle\!\infty} V_2}\Big)\epsilon^2m_{\scriptscriptstyle\!\infty} re^{-2m_{\scriptscriptstyle\!\infty} r}
\\&+\frac{5 V_1}{4M_\text{Pl}m^2_{\scriptscriptstyle\!\infty}}\epsilon (m_{\scriptscriptstyle\!\infty} r)^2\text{Ei}(-m_{\scriptscriptstyle\!\infty} r)-\Big(\frac{V_2}{m^2_{\scriptscriptstyle\!\infty}}-\frac{5V^2_1}{2M^2_\text{Pl}m^4_{\scriptscriptstyle\!\infty}}\Big)\epsilon^2(m_{\scriptscriptstyle\!\infty} r)^2\text{Ei}(-2m_{\scriptscriptstyle\!\infty} r)
\bigg]~.
\end{split}
\end{align}
\end{subequations}
\end{widetext}
Note that, the Jordan frame metrics contain the form of a Yukawa potential, which is controlled by the screened parameter. 

\subsection{PPN parameters $\gamma$, $\beta$  and effective gravitational constant $G_{\rm eff}$}
Now, let us calculate the PPN parameters $\gamma$, $\beta$ and the effective gravitational constant $G_{\rm eff}$ as given in the Jordan frame metric \eqref{hJ}. In this subsection, we neglect the cosmological constant $V_{\scriptscriptstyle\rm \!V\!E\!V}$ in the metric, since its effect is very weak on solar system scales. In next subsection, we will discuss its effect separately on cosmological scales.

Using the relations \eqref{t-chi-J} and \eqref{MJE}, from the relations \eqref{JF metric component} and \eqref{hJ} we can identify the PPN parameters $\gamma(r,\epsilon) $, $\beta(r,\epsilon)$ and the effective gravitational constant $G_{\rm eff}(r,\epsilon)$ in the following form,
\begin{widetext}\label{gamma beta}
\begin{subequations}
\begin{align}
\gamma(r,\epsilon)\label{gamma}
=&1-\frac{\Big(\frac{2A_1M_\text {Pl}}{A_{\scriptscriptstyle\rm \!V\!E\!V}}-\frac{3V_1}{2M_\text {Pl}m^2_{\scriptscriptstyle\!\infty}}\Big)\epsilon e^{-m_{\scriptscriptstyle\!\infty} r}}{1+\Big(\frac{A_1M_\text {Pl}}{A_{\scriptscriptstyle\rm \!V\!E\!V}}-\frac{V_1}{M_\text {Pl}m^2_{\scriptscriptstyle\!\infty}}\Big)\epsilon e^{-m_{\scriptscriptstyle\!\infty} r}},
\\
\begin{split}
\beta(r,\epsilon)\label{beta}
=&1-\! \frac{1}{\bigg[\!1\!+\!\Big(\!\frac{A_1M_\text {Pl}}{A_{\scriptscriptstyle\rm \!V\!E\!V}}\!-\!\frac{V_1}{M_\text {Pl}m^2_{\scriptscriptstyle\!\infty}}\!\Big)\!\epsilon e^{-m_{\scriptscriptstyle\!\infty} \!r}\!\bigg]^2}
\!\bigg\{\!\!-\!\frac{3V_1}{4M_\text {\!Pl}m^2_{\scriptscriptstyle\!\infty}}\!\Big(\!1\!+\!\frac53m_{\scriptscriptstyle\!\infty}r\!\Big)\!\epsilon e^{-m_{\scriptscriptstyle\!\infty}\! r}\!\!+\!M^2_\text{\!Pl}\!\Big(\!\frac{A^2_1}{2A^2_{\scriptscriptstyle\rm \!V\!E\!V}}\!-\!\frac{A_2}{A_{\scriptscriptstyle\rm \!V\!E\!V}}\!\Big)\!\epsilon^2e^{-2m_{\scriptscriptstyle\!\infty}\! r}
\\&~+\frac{3V^2_1}{4M^2_\text {Pl}m^4_{\scriptscriptstyle\!\infty}}\Big(1+\frac13 m_{\scriptscriptstyle\!\infty} r\Big)\epsilon^2e^{-2m_{\scriptscriptstyle\!\infty} r}+\frac{V_2}{2m^2_{\scriptscriptstyle\!\infty}}\Big(1-\frac{3V^2_1}{M^2_{\text {Pl}}m^2_{\scriptscriptstyle\!\infty} V_2}\Big)\epsilon^2(m_{\scriptscriptstyle\!\infty} r)e^{-2m_{\scriptscriptstyle\!\infty} r}
\\&~-\frac{5 V_1}{4M_\text{Pl}m^2_{\scriptscriptstyle\!\infty}}\epsilon \,(m_{\scriptscriptstyle\!\infty}r)^2\,\text{Ei}(-m_{\scriptscriptstyle\!\infty} r)+\Big(\frac{V_2}{m^2_{\scriptscriptstyle\!\infty}}-\frac{5V^2_1}{2M^2_\text{Pl}m^4_{\scriptscriptstyle\!\infty}}\Big)\epsilon^2(m_{\scriptscriptstyle\!\infty}r)^2\text{Ei}(-2m_{\scriptscriptstyle\!\infty} r)
\bigg\},
\end{split}
\\\begin{split}\label{Geff}G_{\!\rm eff}(r,\epsilon)\!=&GA^2_{\scriptscriptstyle\rm \!V\!E\!V}\bigg[1+\Big(\frac{A_1M_\text {Pl}}{A_{\scriptscriptstyle\rm \!V\!E\!V}}-\frac{V_1}{M_\text {Pl}m^2_{\scriptscriptstyle\!\infty}}\Big)\epsilon e^{-m_{\scriptscriptstyle\!\infty} r}\bigg].\end{split}
\end{align}
\end{subequations}
\end{widetext}
This is one of the main results of this article. The Taylor coefficients $(V_{\scriptscriptstyle\rm \!V\!E\!V},\,V_1,\,V_2;\,A_{\scriptscriptstyle\rm \!V\!E\!V},\,A_1,\,A_2)$, the screened parameter $\epsilon$, and the effective mass $m_{\scriptscriptstyle\!\infty}$, can all be obtained from two arbitrary functions $V(\phi)$ and $A(\phi)$. Obviously, the PPN parameters and the effective gravitational constant depend not only on the distance $r$ between the source object and the test mass, but also on the screened parameter $\epsilon$. The screened parameter depends on background matter density $\rho_{\scriptscriptstyle\!\infty}$ and the physical properties (density  $\rho_0$ and radius $R$) of the source object. That is to say, there are different PPN parameters and effective gravitational constants for different sources in SMG theories. Therefore, the observational constraints in the solar system, including the Cassini constraint and the perihelion shift of Mercury constraint, etc., are applicable only to the Sun but not to other sources in SMG theories.

Note that, for the compact objects (such as the Sun, the Earth and the Moon), the screening effect is very strong and the fifth force is much weaker than the gravitational force. However, for galaxies and galaxy clusters, their densities are very low, the screening effect becomes weak and the fifth force becomes comparable with the gravitational force. The extra fifth force may manifestly change the behaviour of the circular velocity for the test objects in the outskirts of galactic halo \cite{chameleon-galaxy} {\tc{and be involved to explain their observed cored density distribution \cite{donato}}. The scalar field may be screened in the interior of the cluster, while its outer region can still be affected by the fifth force. The potential governing the dynamics of the matter fields can differ significantly from the lensing potential, which leads to a difference between the mass of the halo obtained from dynamical measurements (e.g., velocity dispersion) and that obtained from gravitational lensing \cite{symmetron-galaxy,SMG-cluster}. So, we expect that the model parameter space of SMG would be further depressed if observations at
galactic scales were included. This issue will be addressed in our future study.

In the solar system, the distance $r$ is always much less than the Compton wavelength $m^{-1}_{\scriptscriptstyle\!\infty}$, which roughly is cosmological scales, i.e., $m_{\scriptscriptstyle\!\infty} r\ll1$ is satisfied. At the same time, the screening effect is very strong for the Sun (dense body) and the screened parameter $\epsilon\ll1$. In the case of $x\ll1$, the asymptotic behavior of the exponential integral function ${\rm Ei}(-x)$ is
\be
{\rm Ei}(-x)\simeq\ln x+\bm\gamma_{\scriptscriptstyle \!E\!M}-x+\frac{x^2}{4}+\mathcal{O}(x^3),
\ee
where $\bm\gamma_{\scriptscriptstyle \!E\!M}\!=\! 0.57721\!\cdots$ is the Euler-Mascheroni constant. Therefore, in the case $m_{\scriptscriptstyle\!\infty} r\ll1$, the terms involving ${(m_{\scriptscriptstyle\!\infty} r)}^2{\rm Ei}(-m_{\scriptscriptstyle\!\infty} r)$ fall off proportional to  $ {(m_{\scriptscriptstyle\!\infty} r)}^2\ln(-m_{\scriptscriptstyle\!\infty} r)$, and the terms involving ${(m_{\scriptscriptstyle\!\infty} r)}\exp(-m_{\scriptscriptstyle\!\infty} r)$ fall off proportional to ${m_{\scriptscriptstyle\!\infty} r}$. All these terms may be neglected. Thus, the PPN parameters and the effective gravitational constant are simplified as
\begin{subequations}\label{gamma,beta-epsilon}
\begin{align}
\begin{split}
\gamma(\epsilon)\label{gamma-epsilon}
=&1-\bigg(\frac{2A_1M_\text {Pl}}{A_{\scriptscriptstyle\rm \!V\!E\!V}}-\frac{3V_1}{2M_\text {Pl}m^2_{\scriptscriptstyle\!\infty}}\bigg)\epsilon
\\&\,\,\,
+\!\bigg(\!\frac{2A^2_1M^2_\text {Pl}}{A^2_{\scriptscriptstyle\rm \!V\!E\!V}}\!-\!\frac{7V_1A_1}{2m^2_{\scriptscriptstyle\!\infty}A_{\scriptscriptstyle\rm \!V\!E\!V}}\!+\!\frac{3V^2_1}{2M^2_\text {Pl}m^4_{\scriptscriptstyle\!\infty}}\!\bigg)\epsilon^2,
\end{split}
\\
\begin{split}
\beta(\epsilon)\label{beta-epsilon}
=&1+\frac{3V_1}{4M_\text {Pl}m^2_{\scriptscriptstyle\!\infty}}\epsilon+\!\bigg[\Big(\frac{A_2}{A_{\scriptscriptstyle\rm \!V\!E\!V}}-\frac{A^2_1}{2A^2_{\scriptscriptstyle\rm \!V\!E\!V}}\Big)M^2_\text {Pl}
\\&\,\,\,
-\frac{3V_1A_1}{2m^2_{\scriptscriptstyle\!\infty}A_{\scriptscriptstyle\rm \!V\!E\!V}}+\frac{3V^2_1}{4M^2_\text {Pl}m^4_{\scriptscriptstyle\!\infty}}\bigg]\epsilon^2,
\end{split}
\\
\begin{split}\label{GJr0epsilon0}
G_{\!\rm eff}(\epsilon)=&GA^2_{\scriptscriptstyle\rm \!V\!E\!V}\bigg[1+\Big(\frac{A_1M_\text {Pl}}{A_{\scriptscriptstyle\rm \!V\!E\!V}}-\frac{V_1}{M_\text {Pl}m^2_{\scriptscriptstyle\!\infty}}\Big)\epsilon\bigg].
\end{split}
\end{align}
\end{subequations}
%
These relations are applicable to the solar system (or other solar systems), in which the screening effect is very strong $\epsilon\ll1$, and the PPN parameters $\gamma$ and $\beta$ are both close to unity. Comparing the effective gravitational constant \eqref{GJr0epsilon0} with the PPN parameter $\gamma$ \eqref{gamma-epsilon}, we find the approximate relation
\begin{align}\label{G and GJ}
\begin{split}
G_{\!\rm eff}(\epsilon)\simeq&GA^2_{\scriptscriptstyle\rm \!V\!E\!V}\Big[1-\frac{\gamma(\epsilon)-1}{2}\Big].
\end{split}
\end{align}
In fact, in the case $\epsilon\ll1$, a general relation like this can be obtained from the relations \eqref{Geff} and \eqref{gamma},
\begin{align}\label{G and GJ-r}
\begin{split}
G_{\!\rm eff}(r,\epsilon)\simeq&GA^2_{\scriptscriptstyle\rm \!V\!E\!V}\Big[1-\frac{\gamma(r,\epsilon)-1}{2}\Big],
\end{split}
\end{align}
which is applicable to the generic SMG.

Let us consider a general coupling function $A(\phi)$ in the form,
\be
A(\phi)=1+\sum_{n=1}^{+\infty}a_n\Big(\frac{\phi-\phi_\star}{M_{\rm Pl}}\Big)^n,
\ee
where $a_n$ and $\phi_\star$ are free parameters. Using the screened parameter \eqref{epsilon}, the coupling function VEV can be expressed as
\be\label{Avev}
A_{\scriptscriptstyle\rm \!V\!E\!V}\sim 1+ \sum_{n=1}^{+\infty}a_n\big(\Phi_{\rm E}\,\epsilon\big)^n,
\ee
where $\Phi_{\rm E}$ is the Newtonian potential at the surface of the source object in the Einstein frame. For the compact objects (such as the Sun, the Earth and the Moon), $\Phi_{\rm E}$ is always much less than unity, and the screening effect is very strong $\epsilon\ll1$, which follows that $|A_{\scriptscriptstyle\rm \!V\!E\!V}-1|\ll1$. 
Using this result and the Cassini constraint $|\gamma_{\rm obs}-1|\lesssim 2.3\times10^{-5}$ \cite{nature-gamma}, from the relation \eqref{G and GJ}, we have
\be
\frac{|G_{\!\rm eff}(\epsilon_{\rm Sun})-G|}{G}\simeq\frac{|\gamma_{\rm Sun}-1|}{2}\lesssim 1.1\times10^{-5},
\ee
which is applicable to any generic SMG. This result implies that the effective gravitational constant $G_{\!\rm eff}(\epsilon_{\rm Sun})$ is approximately equal to the Newtonian gravitational constant $G$ within $10^{-5}$ accuracy in the solar system.

For the limiting case with $\epsilon\rightarrow0$, from the relations \eqref{gamma,beta-epsilon} and \eqref{Avev}, we obtain $\gamma\rightarrow1$, $\beta\rightarrow1$, $G_{\rm eff}\rightarrow G$. These imply that SMG converges back to GR in this limiting case, because of the PPN parameters $\gamma=\beta=1$ in GR \cite{Will:1993ns,WillReview}.

\subsection{Effective cosmological constant}\label{cc}
SMG contains a scalar degree of freedom, whose potential can naturally provide the vacuum energy required to drive cosmic acceleration at late times. More precisely, SMG requires that the effective potential of scalar field has a minimum, which can be understood as a stable vacuum. Around this minimum (physical vacuum), the bare potential has a VEV, which can play the role of cosmological constant (or, equivalently, the dark energy). In this subsection, we will discuss this issue for the generic SMG.

Considering the metric of SMG around the dense object (such as white dwarf, neutron star and black hole), the screened parameter is $\epsilon\rightarrow0$.
In this limiting case, from the relation \eqref{Avev}, we have $A_{\scriptscriptstyle\rm \!V\!E\!V}\rightarrow1$. Using this, and the relations in \eqref{metric EF to JF} and \eqref{MJE}, we derive
\begin{subequations}
\begin{align}
g^{\rm J}_{\mu\nu}&\rightarrow g^{\rm E}_{\mu\nu}
\\t_{\rm J}\rightarrow t_{\rm E},\,\,\,\chi&\rightarrow r,\,\,\,M_{\rm J}\rightarrow M_{\rm E},
\end{align}
\end{subequations}
which imply that the Einstein and Jordan frame converge to the same frame in this limit. Furthermore, in this limit, from the Jordan frame metric \eqref{hJ} or the Einstein frame metric \eqref{hE001} and \eqref{hE111}, we find that these two frame metrics both converge to
%
\begin{align}
\begin{split}
ds^2\simeq&-\big( 1-\frac{2GM}{r}-\frac{\Lambda}{3}r^2\big)dt^2
\\&+\big( 1+\frac{2GM}{r}-\frac{\Lambda}{6}r^2\big)\big( dr^2 + r^2 d\Omega^2\big)
\end{split}
\end{align}
with
\be
\Lambda\equiv {8\pi G}V_{\scriptscriptstyle\rm \!V\!E\!V}.
\ee
This is the isotropic form of Schwarzschild-(A)de Sitter metric \cite{f(R)-S-AdS} in the weak field limit. Using the coordinate transformation
\be
r\simeq\tilde{r}\big(1-\frac{GM}{\tilde r}+\frac{\Lambda}{12}{\tilde r}^2\big),
\ee
we obtain the standard form of Schwarzschild-(A)de Sitter metric in the weak field limit,
\begin{align}
\begin{split}
ds^2\simeq&-\big( 1-\frac{2GM}{\tilde r}-\frac{\Lambda}{3}{\tilde r}^2\big)dt^2
\\&+\big(1+\frac{2GM}{\tilde r}+\frac{\Lambda}{3}{\tilde r}^2\big)d{\tilde r}^2 + {\tilde r}^2 d\Omega^2.
\end{split}
\end{align}
It is easy to identify the cosmological constant $\Lambda$, and we can see that SMG converges back to GR with a cosmological constant in the limit $\epsilon\rightarrow0$.
Thus, the density of the effective cosmological constant (or effective `dark energy') is given by
\be\label{V_DE}
\rho_\Lambda=V_{\scriptscriptstyle\rm \!V\!E\!V}=V[\phi_{\scriptscriptstyle\rm \!V\!E\!V}(\rho_{m})],
\ee
which can be constrained by various cosmological observations.
In addition, in order to consist with current observations, the dark energy density should nearly equal to a constant and the evolution with the redshift should be slow, which is beyond the scope of present work. In this paper, we shall only consider the current energy density of the effect `dark energy' (labeled by the subscript `0'), and constrain the parameters of some specific SMG models, including chameleon, symmetron and dilaton.


\section{Solar System and cosmological constraints}\label{section5}
There are different experimental constraints on the PPN parameters $\gamma$ and $\beta$. Currently, the high accuracy experimental constraints mainly come from the solar system tests. The most stringent constraint on $\gamma$ in the solar system comes from the measurements of Cassini spacecraft, which measured the Shapiro time delay of a radio signal sent from and to the Cassini spacecraft while close to conjunction with the Sun, and got $\gamma_{\rm obs}-1=(2.1\pm 2.3)\times10^{-5}$ at the $1\sigma$ confidence level \cite{nature-gamma}.

The most stringent constraint on $\beta$ comes from measurements of the perihelion shift of Mercury, which depends on the combination $|2\gamma - \beta -1|$ of the PPN parameters and the solar quadrupole moment $J_2$. The latest inversions of helioseismology data give $J_2 =(2.2 \pm 0.1) \times 10^{-7}$~\cite{2008A&A...477..657A}. Adopting the Cassini bound on $\gamma$, these analyses yield a bound on $\beta_{\rm obs}-1=(-4.1\pm 7.8)\times10^{-5}$ \cite{WillReview}.

A number of advanced experiments or space missions are under development or have been proposed, which could lead to significant improvements in values of the PPN parameters. The Gaia satellite was launched from Europe's Spaceport in 2013, which is located around the L2 Lagrange point of the Sun-Earth system. The Gaia satellite is a high-precision astrometric orbiting telescope, it could measure light-deflection, and is expected to improve the constraint on $\gamma$ to the $10^{-6}$ level \cite{PIAU}. The BepiColombo is a mission to explore the planet Mercury, which is scheduled for launch in 2017. An eight-year mission could yield further
improvements by factors of 2\,--\,5 in $\beta$ \cite{milani02,2007PhRvD..75b2001A}.


For the cosmological constraints, we need the current values of the dark energy density $\rho_{\Lambda_0}$ and the cosmological matter density $\rho_{m_0}$, or equivalently, the current values of the density parameters $\Omega_{\Lambda_0}$ and $\Omega_{m_0}$ and the Hubble constant $H_0$. The latest results come from the observations of Planck satellite, the best-fit values of these parameters are $\Omega_{\Lambda_0}=0.683,~ \Omega_{m_0}=0.317, ~H_0=67.3~{\rm km\cdot s^{-1}Mpc^{-1}}$ \cite{Planck 2013 Cosmological parameters}.

In this section, we will focus on three specific theories of SMG (chameleon, symmetron and dilaton models). By investigating these models on solar system and cosmological scales, we will derive the combined constraints on model parameters.

\subsection{Chameleons}\label{Chameleons}
\subsubsection{The original chameleon}
In order that a certain massive scalar-tensor gravity can satisfy the solar system experiments, the chameleon model was introduced as a screening mechanism by Khoury and Weltman \cite{chameleon cosmology,chameleon fields,chameleon1}. The original chameleon model is characterized by the Ratra-Peebles runaway potential and an exponential coupling function
\begin{subequations}
\begin{align}
V(\phi)&=\frac{M^{4+\alpha}}{\phi^{\alpha}},
\\
A(\phi)&=\exp\Big(\frac{\xi\phi}{M_\text{Pl}}\Big)\label{A-chameleon},
\end{align}
\end{subequations}
where $M$ is a constant with the dimension of mass, $\xi$ is a positive coupling constant, and $\alpha\sim\mathcal O(1)$ is a positive constant index.

The chameleon effective potential has a minimum. Using the relations in \eqref{phi min} and \eqref{m eff}, we obtain the chameleon field value and the effective mass of the chameleon at this minimum,
\begin{subequations}\label{phi-min m-eff C}
\begin{align}
\phi_{\rm min}(\rho)\label{phi-min for chameleon}
&\simeq\bigg(\frac{{\alpha} M_\text{Pl} M^{4+\alpha}}{\xi\rho}\bigg)^{\frac{1}{\alpha+1}},
\\
m^2_{\rm eff}(\rho)\label{m-eff for chameleon}
&\simeq(\alpha+1)\frac{\xi\rho}{M_\text{Pl}\phi_{\rm min}}~.
\end{align}
\end{subequations}
We find that for the higher ambient density $\rho$, the value of $\phi_{\rm min}$ is smaller and the effective mass $m_{\rm eff}$ is larger. The Ratra-Peebles runaway potential $V(\phi)$ and the exponential coupling function $A(\phi)$ are expanded in Taylor's series at the chameleon VEV $\phi_{\scriptscriptstyle\!\infty}\equiv\phi_{\rm min}(\rho_{\scriptscriptstyle\!\infty})$ as follows,
\begin{subequations}
\begin{align}
\begin{split}
V(\phi)
&=\frac{\xi\rho_{\scriptscriptstyle\!\infty}\phi_{\scriptscriptstyle\!\infty}}{\alpha M_\text{Pl}}-\frac{\xi\rho_{\scriptscriptstyle\!\infty}}{M_\text {Pl}}(\phi-\phi_{\scriptscriptstyle\!\infty})
\\&\quad+\frac{(\alpha+1)\xi\rho_{\scriptscriptstyle\!\infty}}{2M_\text{Pl}\phi_{\scriptscriptstyle\!\infty}}(\phi-\phi_{\scriptscriptstyle\!\infty})^2+\cdots
\end{split}
\\
\begin{split}
A(\phi)
&=e^{\frac{\xi\phi_{\scriptscriptstyle\!\infty}}{M_\text{Pl}}}+\frac{\xi}{M_\text{Pl}}e^{\frac{\xi\phi_{\scriptscriptstyle\!\infty}}{M_\text{Pl}}}(\phi-\phi_{\scriptscriptstyle\!\infty})
\\&\quad+\frac{\xi^2}{2M^2_\text{Pl}}e^{\frac{\xi\phi_{\scriptscriptstyle\!\infty}}{M_\text{Pl}}}(\phi-\phi_{\scriptscriptstyle\!\infty})^2+\cdots~.
\end{split}
\end{align}
\end{subequations}
From these formulae, we obtain the expansion coefficients
\begin{subequations}
\begin{align}
\begin{split}\label{V012}
V_{\scriptscriptstyle\rm \!V\!E\!V}=\frac{\xi\rho_{\scriptscriptstyle\!\infty}\phi_{\scriptscriptstyle\!\infty}}{\alpha M_\text{Pl}},~
V_1=-\frac{\xi\rho_{\scriptscriptstyle\!\infty}}{M_\text{Pl}},~
V_2=\frac{(\alpha+1)\xi\rho_{\scriptscriptstyle\!\infty}}{2M_\text{Pl}\phi_{\scriptscriptstyle\!\infty}},
\end{split}
\\
\begin{split}\label{A012}
A_{\scriptscriptstyle\rm \!V\!E\!V}=e^{\frac{\xi\phi_{\scriptscriptstyle\!\infty}}{M_\text{Pl}}},~
A_1=\frac{\xi e^{\frac{\xi\phi_{\scriptscriptstyle\!\infty}}{M_\text{Pl}}}}{M_\text{Pl}},~
A_2=\frac{\xi^2 e^{\frac{\xi\phi_{\scriptscriptstyle\!\infty}}{M_\text{Pl}}}}{2M^2_\text{Pl}},
\end{split}
\end{align}
\end{subequations}
where $\rho_{\scriptscriptstyle\!\infty}$ is the background matter density of the solar system. If considering the cosmological background, $\rho_{\scriptscriptstyle\!\infty}$ is the cosmological matter density $\rho_{m_0}$. However, if considering the galactic background, $\rho_{\scriptscriptstyle\!\infty}$ is the galactic matter density $\rho_{gal}\simeq10^{5}\rho_{m_0}$.

Using these coefficients and the relations in \eqref{phi-min m-eff C} and \eqref{epsilon}, from the relations in \eqref{gamma,beta-epsilon} we obtain the expressions of parameters $(\gamma,\beta,G_{\rm eff})$ as below,
\begin{subequations}\label{chameleon-gamma-beta-G}
\begin{align}
\gamma-1&=-\frac{2\xi\phi_{\scriptscriptstyle\!\infty}}{M_\text{Pl}\Phi},\label{chameleon-gamma}
\\
\begin{split}
\beta-1&=-\frac{3}{4(\alpha+1)}\bigg(\frac{\phi_{\scriptscriptstyle\!\infty}}{M_\text{Pl}}\bigg)^2\frac{1}{\Phi},
\end{split}
\\
\frac{G_{\rm eff}}{G}\!-\!1\!&=\frac{\xi\phi_{\scriptscriptstyle\!\infty}}{M_\text{Pl}\Phi}~,
\end{align}
\end{subequations}
where $\Phi$ is the Newtonian potential at the surface of the source object, and for the Sun we have $\Phi\simeq 2.12\times10^{-6}$. These results are consistent with the previous ones in the literature \cite{Combined-cc-ss}, where only $\gamma$ parameter was obtained. From these formulae, we can also get the relations between these parameters,
\begin{align}\label{chameleon-gamma-beta-G1}
\begin{split}
\beta-1&=-\frac{3\Phi(\gamma-1)^2}{16\xi^2(\alpha+1)}
\\
\frac{G_{\rm eff}}{G}-1&=-\frac{\gamma-1}{2}
\\
A_{\scriptscriptstyle\rm \!V\!E\!V}-1&=-\frac{\Phi(\gamma-1)}{2}.
\end{split}
\end{align}
%
Obviously, $|\beta-1|\ll|\gamma-1|$. Using the Cassini constraint $|\gamma_{\rm obs}-1|\lesssim 2.3\times10^{-5}$, we obtain the constraint on the model parameters,
\begin{align}\label{phi-infty for chameleon}
\frac{\xi\phi_{\scriptscriptstyle\!\infty}}{M_\text{Pl}}\!=\!\xi\!\bigg(\frac{{\alpha} M^{4+\alpha}}{\xi M^{\alpha}_\text{Pl} \rho_{\scriptscriptstyle\!\infty}}\bigg)^{\!\!\frac{1}{\alpha+1}}\!\!&\lesssim 2.4\!\times\!10^{-\!11}.
\end{align}
In addition, the bounds on the other parameters are also derived,
\begin{align}\label{beta-G-A-chameleon}
\begin{split}
|\beta-1|&\lesssim10^{-\!16}\,\,\qquad{\rm for}\,\,\,\xi\!\!\sim\!\mathcal \!O(1)
\\
\Big|\frac{G_{\rm eff}}{G}-1\Big|&\lesssim\!1.1\times\!10^{-\!5}
\\
|A_{\scriptscriptstyle\rm \!V\!E\!V}-1|&\lesssim \!2.4\times\!10^{-\!11},
\end{split}
\end{align}
%
 which strongly indicate that the PPN parameter $\beta=1$, the effective gravitational constant $G_{\rm eff}\simeq G$, and the exponential coupling function VEV $A_{\scriptscriptstyle\rm \!V\!E\!V}=1$ for chameleon.

{Unfortunately, for the original chameleon, it is impossible to explain cosmic acceleration and to pass the solar system experiments at the same time. For the current universe, the cosmological observations give the density ratio $\rho_{\Lambda_0}/\rho_{m_0}=2.15$. However, in the theoretical side, from the relations $V_{\scriptscriptstyle\rm \!V\!E\!V}$ \eqref{V012} and \eqref{V_DE} we get the ratio between them,
\be\label{dark energy to matter density ratio}
\frac{\rho_{\Lambda_0}}{\rho_{m_0}}=\frac{\xi\phi_{\scriptscriptstyle\!\infty}\!(\rho_{m_0})}{\alpha M_\text{Pl}}=2.15,
\ee
where the density $\rho_{\scriptscriptstyle\!\infty}=\rho_{m_0}$, corresponding to the cosmological matter density. Using the relation in \eqref{phi-min for chameleon}, Eq. \eqref{dark energy to matter density ratio} turns into
\begin{equation}  \label{lambdaAlpha0}
\log \!M\!=\!\frac{\alpha\log{m_\text{Pl}}+\log\rho_{\!\Lambda_0}}{4+\alpha}+\frac{\alpha}{4+\alpha}\!\log\!\frac{\alpha\rho_{\!\Lambda_0}}{\sqrt{8\pi}\xi\rho_{m_0}},
\end{equation}
where $m_\text{Pl}\simeq1.22\times10^{19}\rm GeV$ is the Planck mass, and $\rho_{\Lambda_0}\simeq2.51\times10^{-47}\rm GeV^4$ is the dark energy density. In the case with $\xi\sim\mathcal O(1)$ and $\alpha\sim\mathcal O(1)$, the relation \eqref{lambdaAlpha0} is reduced to
\begin{equation}  \label{lambdaAlpha}
\log M(\rm GeV)\simeq \frac{19\alpha-47}{4+\alpha},
\end{equation}
which is the same relation as found in \cite{Combined-cc-ss,stumwshr2007}. This implies that the influences of the coupling constant $\xi$ and the cosmological matter density $\rho_{m_0}$ are much weaker than that of parameter $\alpha$.

From the solar system constraint \eqref{phi-infty for chameleon}, we obtain its equivalent form
\begin{align}\label{solar system-Cassini constraint}
\frac{\xi\phi_{\scriptscriptstyle\!\infty}\!(\rho_{m_0})}{\alpha M_\text{Pl}}\lesssim 2.4\times10^{-11}\!\cdot\!\frac{1}{\alpha}\!\bigg(\frac{\rho_{\scriptscriptstyle\!\infty}}{\rho_{m_0}}\bigg)^{\!\!\frac{1}{\alpha+1}}.
\end{align}
Obviously, in the cases with either the cosmological background ($\rho_{\scriptscriptstyle\!\infty}=\rho_{m_0}$) or the Milky Way galaxy background ($\rho_{\scriptscriptstyle\!\infty}=\rho_{gal}$), the solar system constraint \eqref{solar system-Cassini constraint} is always incompatible with the cosmological relation \eqref{dark energy to matter density ratio} for $\alpha\sim\mathcal O(1)$. In other words, the original chameleon cannot explain cosmic acceleration and pass solar system constraints at the same time, which is consistent with conclusion found in \cite{Combined-cc-ss}.}

\subsubsection{The exponential chameleon}
The original chameleon is ruled out by the combined constraints of the solar system and cosmology. However, the idea of chameleon can be resurrected by modifying the potential in the form,
\begin{align}\label{V-exponential chameleon}
V(\phi)&=M^4\exp\Big(\frac{M^{\alpha}}{\phi^{\alpha}}\Big).
\end{align}
This chameleon model is called the exponential chameleon, and proposed in \cite{modified chamelons}.

We consider the case with
${\phi}/{M}\gg1$. Using the relation \eqref{phi-min for chameleon},
and considering the cosmological matter density $\rho\!=\!\rho_{m_0}\!\simeq1.17\times\!10^{-47}{\rm GeV^4}$, we get that,
\be\label{MC-supposition}
M\gg1.69\times10^{-13}{\rm eV}.
\ee
In this case, the exponential potential \eqref{V-exponential chameleon} is reduced to
\be
V(\phi)= M^4+\frac{M^{4+\alpha}}{\phi^{\alpha}},
\ee
which is equivalent to the Ratra-Peebles runaway potential plus a cosmological constant \eqref{MC0.002}. Therefore, all calculations of the exponential chameleon are the same as the calculations of the original chameleon, except for the effective dark energy density. The dark energy density of the exponential chameleon is given by
\be\label{exp chameleon DE}
\rho_{\Lambda_0}=M^4+\frac{\xi\phi_{\scriptscriptstyle\!\infty}\!(\rho_{m_0})}{\alpha M_\text{Pl}}\rho_{m_0}.
\ee
Taking into account the solar system constraint \eqref{solar system-Cassini constraint}, the cosmological relation \eqref{exp chameleon DE} is simplified to
\be\label{MC0.002}
M= \rho^{1/4}_{\Lambda_0}\simeq 0.002~{\rm eV},
\ee
which is consistent with the relation \eqref{MC-supposition}.
Using this, the solar system constraint on the parameters $\xi$ and $\alpha$ becomes
\be\label{xia-alpha}
\big(19.5-\log\xi\big)\alpha-\log\alpha\gtrsim10.6-\log\frac{\rho_{\scriptscriptstyle\!\infty}}{\rho_{\Lambda_0}}.
\ee
%

\begin{figure}
\begin{center}
\includegraphics[width=8.5cm]{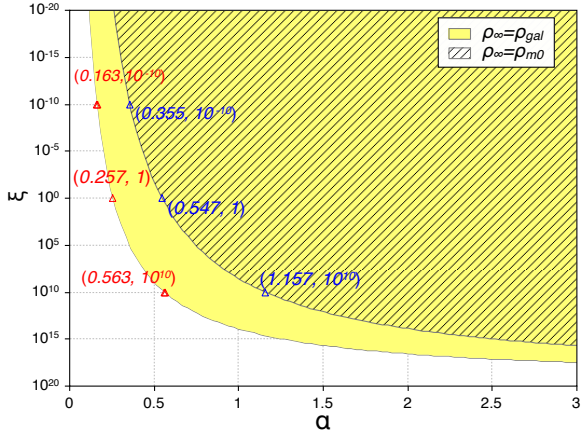}
\caption{\label{xi-alpha} In the parameter space of exponential chameleon models, the shadow region is allowed by Cassini experiment, if assuming the cosmological background, i.e. $\rho_{\scriptscriptstyle\!\infty}=\rho_{m_0}$. While the yellow region is allowed if assuming the galactic background, i.e. $\rho_{\scriptscriptstyle\!\infty}=\rho_{gal}$.}
\end{center}
\end{figure}

In Fig. \ref{xi-alpha}, we plot the constraints on the model parameters $\alpha$ and $\xi$ by considering the cosmological background or the galactic background. In both cases, we find that the constraint on $\xi$ is much looser than that on $\alpha$. For the strong coupling with $\xi\gtrsim1$, we have $\alpha\gtrsim0.547 $ in the case with cosmological background, and $\alpha\gtrsim0.257 $ in the case with galactic background. Even in the limit case with $\xi\gtrsim10^{-10}$, the constraint on $\alpha$ is slightly looser, which is $\alpha\gtrsim0.355$ in the case with cosmological background and $\alpha\gtrsim0.163$ in the case with galactic background.


\subsection{Symmetron}
The symmetron models are characterized by a $\mathbb Z_2$ symmetry breaking potential (a mexican hat potential) and a quadratic coupling function \cite{symmetron-Khoury,symmetron cosmolgy,symmetron-Brax,symmetron-STRUCTURE,symmetron-galaxy},
\begin{subequations}
\begin{align}
V(\phi)&=\mathbb V_0-\frac12\mu^2\phi^2+\frac{\lambda}{4}\phi^4,
\\
A(\phi)&=1+\frac{\phi^2}{2M^2},
\end{align}
\end{subequations}
where $\mu$ and $M$ are mass scales,  $\lambda$ is a positive dimensionless coupling constant, $\mathbb V_0$ is the vacuum energy of the bare potential $V(\phi)$. The effective potential $V_{\rm eff}$ of symmetron has a minimum. Using the relations \eqref{phi min} and \eqref{m eff}, we obtain the field value and the effective mass of the symmetron at this minimum,
\begin{subequations}\label{phi-min m-eff S}
\begin{align}
\phi_{\rm min}(\rho)
&= \left\{
\begin{matrix}
\qquad\quad0\qquad\qquad~~~\,{\rm for}~~~ \rho>\rho_{\rm \scriptscriptstyle SSB}\cr
\pm\frac{\mu}{\sqrt\lambda}\Big(1-\frac{\rho}{\rho_{\rm \scriptscriptstyle SSB}}\Big)^{\frac12}\quad \,{\rm for}~~~ \rho<\rho_{\rm \scriptscriptstyle SSB}\label{phi-min-S}
\end{matrix}
\right.
\\
m^2_{\rm eff}(\rho)
&= \left\{
\begin{matrix}
\mu^2\Big(\frac{\rho}{\rho_{\rm \scriptscriptstyle SSB}}-1\Big)\qquad ~~{\rm for}~~~ \rho>\rho_{\rm \scriptscriptstyle SSB}\cr
2\mu^2\Big(1-\frac{\rho}{\rho_{\rm \scriptscriptstyle SSB}}\Big)\quad ~~~~{\rm for}~~~ \rho<\rho_{\rm \scriptscriptstyle SSB}
\end{matrix}
\right.
\end{align}
\end{subequations}
where $\rho_{\rm \scriptscriptstyle SSB}\equiv M^2\mu^2$ is the critical matter density of spontaneous symmetry breaking (SSB). In high density regions, where $\rho>\rho_{\rm \scriptscriptstyle SSB}$, the effective potential has a minimum at $\phi_{\rm min}=0$, and the $\mathbb Z_2$ symmetry $\phi\rightarrow-\phi$ is ensured. However, in low density regions, where $\rho<\rho_{\rm \scriptscriptstyle SSB}$, the $\mathbb Z_2$ symmetry $\phi\rightarrow-\phi$ is spontaneously broken. In this case, the effective potential has two same minima, and the field settles at one of them. Note that, for either the positive VEV of scalar field or the negative one, the physical results are same, since the scalar field VEV always exists as its square form in the PPN parameters and effective gravitational constant (see Eq. \eqref{symmetron-gamma-beta-G}).
Without loss of generality, we choose the positive scalar field VEV
\be\phi_{\scriptscriptstyle\!\infty}=\frac{\mu}{\sqrt\lambda}\Big(1-\frac{\rho_{\scriptscriptstyle\!\infty}}{\rho_{\rm \scriptscriptstyle SSB}}\Big)^{\frac12}\qquad ~{\rm for}~~~ \rho_{\scriptscriptstyle\!\infty}<\rho_{\rm \scriptscriptstyle SSB}\label{phiVEV-symmetron},\ee
where $\rho_{\scriptscriptstyle\!\infty}$ is the background matter density of the solar system. Similar to the chameleon models, we have $\rho_{\scriptscriptstyle\!\infty}=\rho_{m_0}$ if considering the cosmological matter density as background, while $\rho_{\scriptscriptstyle\!\infty}=\rho_{gal}$ if setting the galactic matter density as background.


The $\mathbb Z_2$ symmetry breaking potential $V(\phi)$ and the quadratic coupling function $A(\phi)$ are expanded in Taylor's series  at this VEV,
\begin{subequations}
\begin{align}
\begin{split}
V(\phi)
&=\mathbb V_0-\frac{\rho^2_{\rm \scriptscriptstyle SSB}-\rho^2_{\scriptscriptstyle\!\infty}}{4\lambda M^4}-\frac{\rho_{\scriptscriptstyle\!\infty}\phi_{\scriptscriptstyle\!\infty}}{M^2}(\phi-\phi_{\scriptscriptstyle\!\infty})
\\&\quad+\Big(\mu^2-\frac{3\rho_{\scriptscriptstyle\!\infty}}{2M^2}\Big)(\phi-\phi_{\scriptscriptstyle\!\infty})^2+\cdots,
\end{split}
\\
\begin{split}
A(\phi)
&=1\!+\!\frac{\phi^2_{\scriptscriptstyle\!\infty}}{2M^2}\!+\!\frac{\phi_{\scriptscriptstyle\!\infty}}{M^2}(\!\phi\!-\!\phi_{\scriptscriptstyle\!\infty}\!)\!+\!\frac{1}{2M^2}\!(\!\phi\!-\!\phi_{\scriptscriptstyle\!\infty}\!)^2\!+\!\cdots.
\end{split}
\end{align}
\end{subequations}
The expansion coefficients are obtained directly,
\begin{subequations}
\begin{align}
\begin{split}
V_{\scriptscriptstyle\rm \!V\!E\!V}\label{Veve-symmetron}
=\mathbb V_0-\frac{\rho^2_{\rm \scriptscriptstyle SSB}-\rho^2_{\scriptscriptstyle\!\infty}}{4\lambda M^4},~V_1=-\frac{\rho_{\scriptscriptstyle\!\infty}\phi_{\scriptscriptstyle\!\infty}}{M^2},~V_2=\mu^2-\frac{3\rho_{\scriptscriptstyle\!\infty}}{2M^2},~
\end{split}
\\
\begin{split}
A_{\scriptscriptstyle\rm \!V\!E\!V}
=1+\frac{\phi^2_{\scriptscriptstyle\!\infty}}{2M^2},~A_1=\frac{\phi_{\scriptscriptstyle\!\infty}}{M^2},~A_2=\frac{1}{2M^2}.
\end{split}
\end{align}
\end{subequations}
%

Using these coefficients and the relations, we get the expressions of $\gamma$, $\beta$ and $G_{\rm eff}$ as follows,
\begin{subequations}\label{symmetron-gamma-beta-G}
\begin{align}
\gamma-1&=-\frac{2\phi^2_{\scriptscriptstyle\!\infty}}{M^2\Phi},\label{gamma-symmetron}
\\
\beta-1&=\frac{1}{2}\bigg(\frac{\phi_{\scriptscriptstyle\!\infty}}{M\Phi}\bigg)^2,
\\
\frac{G_{\rm eff}}{G}-1&=\frac{\phi^2_{\scriptscriptstyle\!\infty}}{M^2\Phi}~.
\end{align}
\end{subequations}
These results are consistent with the previous ones in the literature \cite{symmetron-Khoury}, where only the $\gamma$ parameter was obtained. From these formulae, we can also get the useful relations between these parameters,
\begin{align}\label{symmetron-gamma-beta-G1}
\begin{split}
\gamma-1&=-4\Phi(\beta-1),
\\
\frac{G_{\rm eff}}{G}-1&=-\frac{\gamma-1}{2},
\\
A_{\scriptscriptstyle\rm \!V\!E\!V}-1&=\Phi^2(\beta-1).
\end{split}
\end{align}
%
Obviously, for the symmetron we always have $|\gamma-1|\ll|\beta-1|$, so the constraints on the model parameters are mainly from the measurements of $\beta$, instead of $\gamma$. Using the perihelion shift of Mercury constraint $|\beta_{\rm obs}-1|\lesssim 7.8\times10^{-5}$, and the relations in \eqref{symmetron-gamma-beta-G}, we obtain the following bound on the symmetron parameters,
\begin{align}\label{phi-infty-symmetron}
\frac{\rho_{\rm \scriptscriptstyle SSB}-\rho_{\scriptscriptstyle\!\infty}}{\lambda M^4}
&\lesssim 7.0\times10^{-16}.
\end{align}
Using this bound, we also get the constraints on $\gamma$, $G_{\rm eff}$ and $A_{\rm VEV}$ for the symmetron models,
\begin{align}\label{gamma-G-A-symmetron}
\begin{split}
|\gamma-1|&\lesssim 6.6\times10^{-10},
\\
\Big|\frac{G_{\rm eff}}{G}-1\Big|&\lesssim3.3\times10^{-10},
\\
|A_{\scriptscriptstyle\rm \!V\!E\!V}-1|&\lesssim 3.5\times10^{-16},
\end{split}
\end{align}
which strongly indicate that the PPN parameter $\gamma=1$, the effective gravitational constant $G_{\rm eff}=G$, and the quadratic coupling function VEV $A_{\scriptscriptstyle\rm \!V\!E\!V}=1$ for symmetron.

From the formula $V_{\scriptscriptstyle\rm \!V\!E\!V}$ in \eqref{Veve-symmetron} and the relation in \eqref{V_DE} we obtain the energy density of the effective dark energy in symmetron models,
\begin{align}
\begin{split}
\rho_{\Lambda_0}\label{V-DE-symmetron}
& =\mathbb V_0-\frac{\rho^2_{\rm \scriptscriptstyle SSB}-\rho^2_{m_0}}{4\lambda M^4}
\\&=\mathbb V_0-\frac{\mu^4}{4\lambda}+\frac{\rho^2_{m_0}}{4\lambda M^4}~,
\end{split}
\end{align}
where we have used $\rho_{\scriptscriptstyle\!\infty}=\rho_{m_0}$ for the cosmological background.
In order to get the accelerated expansion of the universe ($\rho_{\Lambda_0}>0$), we need
\be
\mathbb V_0>\frac{\mu^4}{4\lambda}-\frac{\rho^2_{m_0}}{4\lambda M^4}>0,
\ee
i.e. the vacuum energy $\mathbb V_0$ of the bare potential must be positive. Now, let us consider two specific cases of $\mathbb V_0$ function.

{\bf Case 1:}~$\mathbb V_0=0$

This is the original symmetron model, suggested in the literature \cite{symmetron-Khoury,symmetron cosmolgy}. From the relation \eqref{V-DE-symmetron}, we can see that the original symmetron has a negative cosmological constant and cannot drive cosmic acceleration at late times, which is consistent with the conclusion in the previous work \cite{symmetron cosmolgy}.

{\bf Case 2:}~$\mathbb V_0=\mu^4/4\lambda$

This kind of models are proposed in \cite{symmetron-modified}. In this case, the density of symmetron dark energy is
\be\label{rho-symmetron}
\rho_{\Lambda_0} =\frac{\rho^2_{m_0}}{4\lambda M^4},
\ee
and the solar system constraint \eqref{phi-infty-symmetron} becomes
\be\label{Mu}
0<\frac{\rho_{\rm \scriptscriptstyle SSB}-\rho_{\scriptscriptstyle\!\infty}}{\rho_{m_0}}\lesssim 8.1\times10^{-17}.
\ee
From these relations, we evaluate the model parameters of symmetron as follows
\be
\lambda M^4=\frac{\rho_{m_0}}{4}\left(\frac{\rho_{m_0}}{\rho_{\Lambda_0}}\right)\simeq1.4\times10^{-48}{\rm GeV^4},
\ee
and
\begin{equation}\label{Mu-symmetron}
M^2\!\mu^2\!=\! \left\{
\begin{matrix}
\rho_{m_0}\!\simeq\!1.2\!\times\!10^{-47}\,{\rm GeV^4}~~~{\rm for~~ CB} \cr
\rho_{gal}\simeq\!10^{-42}~~~~~~\,{\rm GeV^4}~~~{\rm for~~GB}
\end{matrix}
\right.,
\end{equation}
where $\rho_{\scriptscriptstyle\!\infty}=\rho_{m_0}$ and $\rho_{\scriptscriptstyle\!\infty}=\rho_{gal}$ correspond to the cosmological background (CB) and the galactic background (GB), respectively. 

\subsection{Dilaton}
The dilaton model, inspired by string theory in the large string coupling limit, has an exponentially runaway potential and a quadratic coupling function \cite{String,Dilaton and modified gravity,dilaton-Nonlinear structure formation,dilaton-Morris},
\begin{subequations}
\begin{align}
V(\phi)&=\mathcal V_0\exp\Big(-\frac{\phi}{M_\text{Pl}}\Big),
\\
A(\phi)&=1+\frac{(\phi-\phi_\star)^2}{2M^2},
\end{align}
\end{subequations}
where $\mathcal V_0$ is a constant with the dimension of energy density, $M$ labels the energy scale of the theory, and $\phi_\star$ is approximately the value of $\phi$ today.

The dilaton effective potential $V_{\rm eff}$ also has a minimum. Using the relations \eqref{phi min} and \eqref{m eff}, we obtain the dilaton field value and the effective mass of the dilaton at this minimum,
\begin{subequations}\label{phi-min m-eff D}
\begin{align}
\phi_{\rm min}(\rho)&\simeq\phi_\star+\frac{M^2\mathcal V_0}{M_\text{Pl}\rho}e^{-\frac{\phi_\star}{M_\text{Pl}}},
\\
m^2_{\rm eff}(\rho)&\simeq\frac{\rho}{M^2}+\frac{\mathcal V_0}{M^2_\text{Pl}}e^{-\frac{\phi_\star}{M_\text{Pl}}}.
\end{align}
\end{subequations}
The exponentially runaway potential $V(\phi)$ and the quadratic coupling function $A(\phi)$ can be expanded in Taylor's series at the dilaton VEV $\phi_{\scriptscriptstyle\!\infty}\equiv\phi_{\rm min}(\rho_{\scriptscriptstyle\!\infty})$,
\begin{subequations}
\begin{align}
\begin{split}
V(\phi)
&=\mathcal V_0e^{-\frac{\phi_{\scriptscriptstyle\!\infty}}{M_\text{Pl}}}-\frac{\mathcal V_0}{M_\text{Pl}}e^{-\frac{\phi_{\scriptscriptstyle\!\infty}}{M_\text{Pl}}}(\phi-\phi_{\scriptscriptstyle\!\infty})
\\&\quad+\frac{\mathcal V_0}{2M^2_\text{Pl}}e^{-\frac{\phi_{\scriptscriptstyle\!\infty}}{M_\text{Pl}}}(\phi-\phi_{\scriptscriptstyle\!\infty})^2+\cdots
\end{split}
\\
\begin{split}
A(\phi)
&=1+\frac{(\phi_{\scriptscriptstyle\!\infty}-\phi_\star)^2}{2M^2}+\frac{\phi_{\scriptscriptstyle\!\infty}-\phi_\star}{M^2}(\phi-\phi_{\scriptscriptstyle\!\infty})
\\&\quad+\frac{1}{2M^2}(\phi-\phi_{\scriptscriptstyle\!\infty})^2+\cdots~.
\end{split}
\end{align}
\end{subequations}
So, the expansion coefficients are derived directly,
\begin{subequations}
\begin{align}
\begin{split}
V_{\scriptscriptstyle\rm \!V\!E\!V}=\mathcal V_0e^{-\frac{\phi_{\scriptscriptstyle\!\infty}}{M_\text{Pl}}},~
V_1=-\frac{\mathcal V_0 e^{-\frac{\phi_{\scriptscriptstyle\!\infty}}{M_\text{Pl}}}}{M_\text{Pl}},~
V_2=\frac{\mathcal V_0 e^{-\frac{\phi_{\scriptscriptstyle\!\infty}}{M_\text{Pl}}}}{2M^2_\text{Pl}},~\label{Vvev-dilaton}
\end{split}
\\
\begin{split}
A_{\scriptscriptstyle\rm \!V\!E\!V}=1+\frac{(\phi_{\scriptscriptstyle\!\infty}-\phi_\star)^2}{2M^2},~
A_1=\frac{\phi_{\scriptscriptstyle\!\infty}-\phi_\star}{M^2},~
A_2=\frac{1}{2M^2}~,
\end{split}
\end{align}
\end{subequations}
where $\rho_{\scriptscriptstyle\!\infty}$ is the background matter density of the solar system.


Using these coefficients and the relations in \eqref{phi-min m-eff D}, \eqref{epsilon} and \eqref{gamma,beta-epsilon}, we obtain the PPN parameters and effective gravitational constant,
\begin{subequations}\label{dilaton-gamma-beta-G}
\begin{align}
\gamma-1&=-\frac{2(\phi_{\scriptscriptstyle\!\infty}-\phi_\star)^2}{M^2\Phi},
\\
\beta-1&=\frac{1}{2}\bigg(\frac{\phi_{\scriptscriptstyle\!\infty}-\phi_\star}{M\Phi}\bigg)^2,
\\
\frac{G_{\rm eff}}{G}-1&=\frac{(\phi_{\scriptscriptstyle\!\infty}-\phi_\star)^2}{M^2\Phi}.
\end{align}
\end{subequations}
The useful relations between them are also derived directly
\begin{align}\label{dilaton-gamma-beta-G1}
\begin{split}
\gamma-1&=-4\Phi(\beta-1),
\\
\frac{G_{\rm eff}}{G}-1&=-\frac{\gamma-1}{2},
\\
A_{\scriptscriptstyle\rm \!V\!E\!V}-1&=\Phi^2(\beta-1).
\end{split}
\end{align}
%
Note that, these relations are exactly same with the ones in symmetron model. Therefore, among the solar system tests, the perihelion shift of Mercury constraint $|\beta_{\rm obs}-1|\lesssim 7.8\times10^{-5}$ follows the most stringent constraint on the model parameter, which is
\begin{align}
\frac{M\mathcal V_0}{M_\text{Pl}\rho_{\scriptscriptstyle\!\infty}}e^{-\phi_\star/M_\text{Pl}}&\lesssim 2.6\times10^{-8}.\label{phi-infty-dilaton}
\end{align}
The bounds of the other parameters in the dilaton models are
\begin{align}\label{gamma-G-A-dilaton}
\begin{split}
|\gamma-1|&\lesssim 6.6\times10^{-10},
\\
\Big|\frac{G_{\rm eff}}{G}-1\Big|&\lesssim3.3\times10^{-10},
\\
|A_{\scriptscriptstyle\rm \!V\!E\!V}-1|&\lesssim 3.5\times10^{-16}.
\end{split}
\end{align}

Using the relations in Eqs. \eqref{Vvev-dilaton} and \eqref{V_DE}, we obtain the density of dilaton dark energy,
\begin{align}\label{dilaton dark energy0}
\begin{split}
\rho_{\Lambda_0}&=\mathcal V_0e^{-\phi_{\scriptscriptstyle\!\infty}/M_\text{Pl}}
\\
&\simeq\mathcal V_0e^{-\phi_\star/M_\text{Pl}}\simeq2.51\times10^{-47}{\rm GeV^4},
\end{split}
\end{align}
where $\rho_{\scriptscriptstyle\!\infty}=\rho_{m_0}$ on cosmological scales. This is consistent with the relation found in \cite{Dilaton and modified gravity}. 
Taking into account this relation, the solar system constraint \eqref{phi-infty-dilaton} turns into
\begin{align}\label{MD0}
\frac{M\rho_{\Lambda_0}}{M_\text{Pl}\rho_{\scriptscriptstyle\!\infty}}&\lesssim 2.6\times10^{-8},
\end{align}
that is,
\begin{subequations}\label{MD1}
\begin{align}
\frac{M}{M_\text{Pl}}&\lesssim 1.2\times10^{-8}\qquad{\rm for}\quad    \rho_{\scriptscriptstyle\!\infty}=\rho_{m_0},
\\
\frac{M}{M_\text{Pl}}&\lesssim 1.2\times10^{-3}\qquad{\rm for}\quad    \rho_{\scriptscriptstyle\!\infty}=\rho_{gal},
\end{align}
\end{subequations}
where $\rho_{\scriptscriptstyle\!\infty}=\rho_{m_0}$ and $\rho_{\scriptscriptstyle\!\infty}=\rho_{gal}$ correspond to the cosmological background and galactic background, respectively.

\section{Conclusions}\label{section6}
Screened modified gravity (SMG) is a kind of scalar-tensor theories with screening mechanisms, which can generate screening effect to suppress fifth force and pass the solar system tests. In this paper, we calculated the PPN parameters $\gamma$ and $\beta$ for SMG with a general potential $V$ and coupling function $A$ in the case of a static spherically symmetric source. In addition, we discussed the effective cosmological constant in the generic SMG. These two analyses allow us to constrain the model parameters by combining the observations on solar system and cosmological scales.

The PPN parameters were typically calculated under the assumption of point source surrounded by vacuum \cite{ppn-st-gen-potential,Testing-st-ppn}, but this assumption is generally not appropriate to solve the massive scalar field. In order to overcome this defect and calculate the PPN parameters for the generic SMG, in which the scalar field is always massive, we solved the scalar field in the Einstein frame in the case of an extended source surrounded by a homogeneous background, which is the more realistic case for the source as the Sun or the Earth. Then, we solved the massless metric field in the Einstein frame. By transforming the results to the Jordan frame through a conformal rescaling of the metric, we obtained the PPN parameters $\gamma$, $\beta$ and the effective gravitational constant $G_{\rm eff}$ for the general SMG models.

We found that the parameters ($\gamma$, $\beta$, $G_{\rm eff}$) depend not only on the distance between the source object and the test mass, but also on the screened parameter $\epsilon$, which is determined by the physical properties of the source object. Moreover, SMG contains a scalar degree of freedom, whose effective potential has a minimum (physical vacuum), and the bare potential has a VEV at this minimum. The bare potential VEV can naturally play the role of dark energy to accelerate the expansion of the universe at late times. So, as anticipated, the SMG could not only pass the strict solar tests, but also account for the accelerated expansion of the universe.



We applied our results to three specific cases of SMG theories (chameleon, symmetron and dilaton models), and calculated their PPN parameters and effective cosmological constant, respectively. By investigating the current experiments on solar system and cosmological scales, we derived the combined parameter constraints on these three models, respectively. Consistent with all the previous works, we found the following results for these SMG models: The original chameleon cannot explain cosmic acceleration and pass solar system constraints at the same time, but this difficulty is overcome in the exponential chameleon. The original symmetron $(\mathbb V_0=0)$ has a negative cosmological constant, and cannot drive cosmic acceleration. However, the modified symmetron with $\mathbb V_0=\mu^4/4\lambda$ can realize it. The dilaton is a fine model for both passing solar system tests and accelerating the expansion of the universe in the late stage. For each of these healthy models (the exponential chameleon, the modified symmetron and the dilaton), we obtained the constraints on the model parameters, respectively.


\begin{acknowledgments}

WZ is supported by Project 973 under Grant No. 2012CB821804, by NSFC No. 11173021, 11322324, 11421303 and project of Knowledge Innovation Program of Chinese Academy of Science. YFC is supported in part by the Chinese National Youth Thousand Talents Program, the USTC start-up funding, and NSFC No. 11421303.

\end{acknowledgments}


\baselineskip=12truept

\begin{thebibliography}{99}

\bibitem{Supernova1998}
A.G. Reiss {\it et al}., \href{http://iopscience.iop.org/article/10.1086/300499} {Astron. J. {\bf116}, 1009 (1998)};
\href{http://dx.doi.org/10.1086/300738}{Astron. J. {\bf117}, 707 (1999)};
S. Perlmutter {\it et al}., \href{http://dx.doi.org/10.1086/307221}{ApJ {\bf 517}, 565 (1999)}.


\bibitem{dark energy}
E.J. Copeland, M. Sami, and  S. Tsujikawa, \href{http://dx.doi.org/10.1142/S021827180600942X}{Int. J. Mod. Phys. D {\bf15}, 1753 (2006)};
J.A. Frieman, M.S. Turner, and D. Huterer, \href{http://dx.doi.org/10.1146/annurev.astro.46.060407.145243}{Ann. Rev. Astron. Astrophys. {\bf46}, 385 (2008)};
M. Li , X.D. Li, S. Wang, and Y. Wang, \href{http://dx.doi.org/10.1088/0253-6102/56/3/24}{Commun. Theor. Phys. {\bf56}, 525 (2011)}.


\bibitem{cc}
S. Weinberg, \href{http://dx.doi.org/10.1103/RevModPhys.61.1} {Rev. Mod. Phys. {\bf61}, 1 (1989)}; \href{http://arxiv.org/abs/astro-ph/0005265}{arXiv:astro-ph/0005265v1};
S.M. Carroll, W.H. Press, and  E.L. Turner, \href{http://dx.doi.org/10.1146/annurev.aa.30.090192.002435}{Ann. Rev. Astron. Astrophys. {\bf30}, 499 (1992)};
S.M. Carroll, \href{http://www.livingreviews.org/lrr-2001-1}{Living Rev. Relativity {\bf4}, 1 (2001)};
V. Sahni, A.A. Starobinsky,  \href{http://inspirehep.net/record/499012}{Int. J. Mod. Phys. D {\bf9}, 373 (2000)};
P.J.E. Peebles and  B. Ratra, \href{http://dx.doi.org/10.1103/RevModPhys.75.559}{Rev. Mod. Phys. {\bf75},  559 (2003)};
T. Padmanabhan, \href{http://dx.doi.org/10.1016/S0370-1573(03)00120-0}{Phys. Rept. {\bf380}, 235 (2003)}.


\bibitem{Quintessence}
R.R. Caldwell, R. Dave, and P.J. Steinhardt,
 \href{http://dx.doi.org/10.1103/PhysRevLett.80.1582} {Phys. Rev. Lett. {\bf 80}, 1582 (1998)};
I. Zlatev, L. Wang and P.J. Steinhardt,
 \href{http://dx.doi.org/10.1103/PhysRevLett.82.896} {Phys. Rev. Lett. {\bf82}, 896 (1999)};
P.J. Steinhardt, L. Wang, and I. Zlatev,
\href{http://dx.doi.org/10.1103/PhysRevD.59.123504}{Phys. Rev. D {\bf 59}, 123504 (1999)};
S.M. Carroll,
\href{http://dx.doi.org/10.1103/PhysRevLett.81.3067}{Phys. Rev. Lett. {\bf 81}, 3067 (1998)}.


\bibitem{phantom} R.R. Caldwell, M. Kamionkowski, and N.N. Weinberg, \href{http://dx.doi.org/10.1103/PhysRevLett.91.071301}{Phys. Rev. Lett. {\bf91}, 071301 (2003)};
R.R. Caldwell, \href{http://dx.doi.org/10.1016/S0370-2693(02)02589-3}{Phys. Lett. B {\bf545}, 23 (2002)}.


\bibitem{quintom}
B. Feng, X.L. Wang, and X.M. Zhang, \href{http://dx.doi.org/10.1016/j.physletb.2004.12.071}{Phys. Lett. B {\bf607}, 35 (2005)};
Z. Guo, Y. Piao, X.M. Zhang, and Y.Z. Zhang, \href{http://dx.doi.org/10.1016/j.physletb.2005.01.017}{ Phys. Lett. B {\bf608},177 (2005)};
W. Zhao and Y. Zhang, \href{http://dx.doi.org/10.1103/PhysRevD.73.123509}{Phys. Rev. D {\bf 73}, 123509 (2006)};
Y.F. Cai, E.N. Saridakis,  M.R. Setare, and J.Q. Xia, \href{http://dx.doi.org/10.1016/j.physrep.2010.04.001}{Phys. Rept. {\bf493}, 1 (2010)}.


\bibitem{modified gravity}
S. Nojiri and S. D. Odintsov,
\href{http://dx.doi.org/10.1016/j.physrep.2011.04.001}{Phys. Rept. {\bf505}, 59 (2011)};
T. Clifton, P.G. Ferreira,  A. Padilla, and C. Skordis,
\href{http://dx.doi.org/10.1016/j.physrep.2012.01.001}{Phys. Rept. {\bf513},1 (2012)};
T. P. Sotiriou and V. Faraoni, \href{http://dx.doi.org/10.1103/RevModPhys.82.451}{Rev. Mod. Phys. {\bf82}, 451 (2010)};
S. Capozziello and M. D. Laurentis, \href{http://dx.doi.org/10.1016/j.physrep.2011.09.003}{Phys. Rept. {\bf509}, 167 (2011)};
J. Sakstein, \href{http://arxiv.org/abs/arXiv:1502.04503}{arXiv:1502.04503}.


\bibitem{Boisseau-2000PRD}B.~Boisseau, G.~Esposito-Farese, D.~Polarski, and A.A. Starobinsky, \href{http://dx.doi.org/10.1103/PhysRevLett.85.2236}{ Phys. Rev. Lett.  {\bf 85}, 2236 (2000)}.


\bibitem{weinberg} S.~Weinberg and E.~Witten, \href{http://dx.doi.org/10.1016/0370-2693(80)90212-9}{Phys.~Lett.~B {\bf 96}, 59 (1980)}.


\bibitem{Cho-1975-PRD}Y.M. Cho and P.G.O. Freund, \href{http://dx.doi.org/10.1103/PhysRevD.12.1711}{Phys. Rev. D {\bf12}, 1711 (1975)}.


\bibitem{string}K. Becker, M. Becker, and J.H. Schwarz, \textit{String Theory and M-Theory a Modern Introduction}, (Cambridge University Press, Cambridge, 2006).


\bibitem{higgs}G. Aad {\it et al}. (ATLAS Collaboration), \href{http://dx.doi.org/10.1016/j.physletb.2013.08.026} {Phys. Lett. B {\bf 726}, 120 (2013)}.


\bibitem{inflation.guth} A.H. Guth, \href{http://dx.doi.org/10.1103/PhysRevD.23.347} {Phys. Rev. D {\bf 23}, 347 (1981)}.


\bibitem{inflation-reheating}B.A. Bassett, S. Tsujikawa, and D. Wands, \href{http://dx.doi.org/10.1103/RevModPhys.78.537}{Rev. Mod. Phys. {\bf78},  537 (2006)}.


\bibitem{Scalar-Tensor}
Y. Fujii and K.I. Maeda,\textit{The Scalar-Tensor Theory of Gravitation}, (Cambridge University Press, Cambridge, 2003);
V. Faraoni,\textit{Cosmology in Scalar-Tensor Gravity}, (Kluwer Academic Publishers, Dordrecht, 2004).


\bibitem{scalar-tensor Solar System}
 N.C. Devi, S. Panda, and A.A. Sen,
\href{http://dx.doi.org/10.1103/PhysRevD.84.063521}{Phys. Rev. D {\bf84}, 063521 (2011)};
S. Tsujikawa, K. Uddin, S. Mizuno, R. Tavakol, and J. Yokoyama,
\href{http://dx.doi.org/10.1103/PhysRevD.77.103009}{Phys. Rev. D {\bf 77}, 103009 (2008)};
O. Minazzoli and A. Hees, \href{http://dx.doi.org/10.1103/PhysRevD.88.041504}{Phys. Rev. D {\bf 88}, 041504 (2013)}.


\bibitem{Scalar-tensor cosmologies}
T. Damour and K. Nordtvedt,
\href{http://dx.doi.org/10.1103/PhysRevLett.70.2217}{Phys. Rev. Lett. {\bf 70}, 2217 (1993)};
\href{http://dx.doi.org/10.1103/PhysRevD.48.3436}{Phys. Rev. D {\bf 48}, 3436 (1993)};
L. J\"arv, P. Kuusk, and M. Saal,
\href{http://dx.doi.org/10.1103/PhysRevD.78.083530}{Phys. Rev. D {\bf78}, 083530 (2008)};
\href{http://dx.doi.org/10.1088/1742-6596/283/1/012017}{J. Phys.:Conf. Ser. {\bf283}, 012017 (2011)};
\href{http://dx.doi.org/10.1103/PhysRevD.85.064013}{Phys. Rev. D {\bf85}, 064013 (2012)};
P. Kuusk, L. J\"arv, and M. Saal,
\href{http://www.worldscientific.com/doi/abs/10.1142/S0217751X09045133}{Int. J. Mod. Phys. A {\bf24}, 1631 (2009)};
\href{http://dx.doi.org/10.1088/1742-6596/343/1/012064}{J. Phys.: Conf. Ser. {\bf343}, 012064 (2012)},
T. Chiba and M. Yamaguchi,
\href{http://dx.doi.org/10.1088/1475-7516/2013/10/040}{JCAP {\bf10}, 040 (2013)}.


\bibitem{f(R)}
T. Faulkner, M. Tegmark, E.F. Bunn, and Y. Mao,
\href{http://dx.doi.org/10.1103/PhysRevD.76.063505}{Phys. Rev. D {\bf76}, 063505 (2007)};
M. Capone and M.L. Ruggiero,
\href{http://dx.doi.org/10.1088/0264-9381/27/12/125006}{Class. Quant. Grav. {\bf27}, 125006 (2010)}.


\bibitem{Brans-Dicke-PPN}
L. Perivolaropoulos,
\href{http://dx.doi.org/10.1103/PhysRevD.81.047501}{Phys. Rev. D {\bf 81}, 047501 (2010)};
Kh. Saaidi, A. Mohammadi, and H. Sheikhahmadi,
\href{http://dx.doi.org/10.1103/PhysRevD.83.104019}{Phys. Rev. D {\bf 83}, 104019 (2011)};
J.W. Moffat and V.T. Toth,
\href{http://www.worldscientific.com/doi/abs/10.1142/S0218271812500848}{Int. J. Mod. Phys. D {\bf21}, 1250084 (2012)};
S. Das and N. Banerjee,
\href{http://dx.doi.org/10.1103/PhysRevD.78.043512}{Phys. Rev. D {\bf 78}, 043512 (2008)};
M. Roshan and F. Shojai,
 \href{http://dx.doi.org/10.1088/0264-9381/28/14/145012}{Class. Quant. Grav. {\bf28}, 145012 (2011)};
L. J\"arv, P. Kuusk, M. Saal, and O. Vilson,
\href{http://dx.doi.org/10.1103/PhysRevD.91.024041}{Phys. Rev. D {\bf 91}, 024041 (2015)}.


\bibitem{String}
T. Damour and A.M. Polyakov, \href{http://dx.doi.org/10.1016/0550-3213(94)90143-0}{ Nucl.~Phys.~B {\bf423}, 532 (1994)};
\href{http://link.springer.com/article/10.1007\%2FBF02106709}{Gen. Rel. Grav. {\bf26}, 1171 (1994)}.


\bibitem{Runaway Dilaton}
T. Damour, F. Piazza, and G. Veneziano,
\href{http://dx.doi.org/10.1103/PhysRevD.66.046007}{Phys. Rev. D {\bf66}, 046007 (2002)};
\href{http://dx.doi.org/10.1103/PhysRevLett.89.081601}{Phys. Rev. Lett. {\bf89}, 081601 (2002)};
M. Gasperini, F. Piazza, and G. Veneziano,
\href{http://dx.doi.org/10.1103/PhysRevD.65.023508}{Phys. Rev. D {\bf65}, 023508 (2001)};
L. J\"arv, P. Kuusk, and M. Saal,
\href{http://dx.doi.org/10.1103/PhysRevD.75.023505}{Phys. Rev. D {\bf75}, 023505 (2007)}.


\bibitem{conformal frame}
V. Faraoni and S. Nadeau, \href{http://dx.doi.org/10.1103/PhysRevD.75.023501}{Phys. Rev. D {\bf 75}, 023501 (2007)};
L. J\"arv, P. Kuusk, and M. Saal,
\href{http://dx.doi.org/10.1103/PhysRevD.76.103506}{Phys. Rev. D {\bf 76}, 103506 (2007)}.
M. Postma and M. Volponi,
\href{http://dx.doi.org/10.1103/PhysRevD.90.103516}{Phys. Rev. D {\bf 90}, 103516 (2014)}.


\bibitem{Torsion balance experiments} E.G. Adelberger, J.H. Gundlach, B.R. Heckel, S. Hoedl, and S. Schlamminger, \href{http://dx.doi.org/10.1016/j.ppnp.2008.08.002}{Prog. Part. Nucl. Phys. {\bf62}, 102-134 (2009)}.


\bibitem{fifth force}
J.G. Williams, S.G. Turyshev, and D. Boggs
\href{http://dx.doi.org/10.1088/0264-9381/29/18/184004}{Class. Quant. Grav. {\bf29}, 184004 (2012)};
E. Fischbach and C. Talmadge \href{http://arxiv.org/abs/hep-ph/9606249}{arXiv:hep-ph/9606249}.


\bibitem{chameleon cosmology}
J.~Khoury and A.~Weltman,
\href{http://dx.doi.org/10.1103/PhysRevD.69.044026} {Phys.~Rev.~D {\bf69}, 044026 (2004)}.


\bibitem{chameleon fields}
J.~Khoury and A.~Weltman,
\href{http://dx.doi.org/10.1103/PhysRevLett.93.171104} {Phys. Rev. Lett. {\bf93}, 171104 (2004)}.


\bibitem{chameleon1}
S.S. Gubser and J.~Khoury,
\href{http://dx.doi.org/10.1103/PhysRevD.70.104001} {Phys. Rev. D {\bf 70}, 104001 (2004)}.


\bibitem{modified chamelons}
P.~Brax, C.~van de Bruck and A.C.~Davis,
 \href{http://dx.doi.org/10.1088/1475-7516/2004/11/004}{JCAP {\bf11}, 004 (2004)};
P.~Brax, C.~van de Bruck, A.C.~Davis, J.~Khoury, and A.~Weltman,
\href{http://dx.doi.org/10.1103/PhysRevD.70.123518}{Phys. Rev. D {\bf 70}, 123518 (2004)};
\href{http://dx.doi.org/10.1103/PhysRevD.82.083503}{Phys.~Rev.~D {\bf 82}, 083503 (2010)};
R.~Gannouji, B.~Moraes, D.F.~Mota, D.~Polarski, S.~Tsujikawa, and H.A.~Winther,
\href{http://dx.doi.org/10.1103/PhysRevD.82.124006} {Phys.~Rev.~D {\bf 82}, 124006 (2010)};
A.L. Erickcek, N. Barnaby, C. Burrage, and Z. Huang,
\href{http://dx.doi.org/10.1088/1475-7516/2004/11/004}{Phys. Rev. D {\bf89}, 084074 (2014)};
J. Khoury, \href{http://dx.doi.org/10.1088/0264-9381/30/21/214004}{Class. Quant. Grav. {\bf30}, 214004 (2013)}.


\bibitem{chameleon-galaxy}
R. Pourhasana, N. Afshordib, R.B. Manna, and A.C. Davisc,
\href{http://dx.doi.org/10.1088/1475-7516/2011/12/005}{JCAP {\bf12}, 005 (2011)};
M. Gronke, C. Llinares, D.F. Mota, and H.A. Winther,
\href{http://dx.doi.org/10.1093/mnras/stv496}{MNRAS {\bf449}, 2837 (2015)};
P. Brax and N. Tamanini,
\href{http://dx.doi.org/10.1103/PhysRevD.93.103502}{Phys. Rev. D {\bf93}, 103502 (2016)}.







\bibitem{symmetron-Khoury}
K.~Hinterbichler and J.~Khoury, \href{http://dx.doi.org/10.1103/PhysRevLett.104.231301}{Phys.~Rev.~Lett. {\bf104}, 231301 (2010)}.
\bibitem{symmetron cosmolgy}K.~Hinterbichler, J.~Khoury, A.~Levy, and A.~Matas,
\href{http://dx.doi.org/10.1103/PhysRevD.84.103521}{Phys.~Rev.~D {\bf 84}, 103521 (2011)}.


\bibitem{symmetron-Brax}
P.~Brax, C.~van de Bruck, A.C.~Davis, B.~Li, B.~Schmauch, and D.~J.~Shaw,
\href{http://dx.doi.org/10.1103/PhysRevD.84.123524} {Phys.~Rev.~D {\bf 84},123524 (2011)}.
\bibitem{symmetron-STRUCTURE}
A.C.~Davis, B.~Li, D.F.~Mota, and H.A.~Winther,
\href{http://dx.doi.org/10.1088/0004-637X/748/1/61}{Astrophys. J. {\bf748},  61 (2012)}.


\bibitem{symmetron-galaxy}
H.A. Winther, D. F. Mota, and B. Li,
\href{http://dx.doi.org/10.1088/0004-637X/756/2/166}{Astrophys. J. {\bf756}, 166 (2012)}.




\bibitem{symmetron-modified}
R. Dong, W.H. Kinney and D. Stojkovic,
\href{http://dx.doi.org/10.1088/1475-7516/2014/01/021}{JCAP {\bf1401}, 021 (2014)}.


\bibitem{Dilaton and modified gravity} P.~Brax, C.~van~de~Bruck, A.C.~Davis, and D.J.~Shaw,
\href{http://dx.doi.org/10.1103/PhysRevD.82.063519}{Phys.~Rev.~D {\bf82}, 063519 (2010)}.
\bibitem{dilaton-Nonlinear structure formation}P.~Brax, C.~van~de~Bruck, A.C.~Davis, B.~Li, and D.J.~Shaw,
\href{http://dx.doi.org/10.1103/PhysRevD.83.104026} {Phys.~Rev.~D {\bf83}, 104026 (2011)}.


\bibitem{dilaton-Morris}
J.R. Morris, \href{http://dx.doi.org/10.1007/s10714-011-1204-8}{Gen. Rel. Grav. {\bf43},  2821 (2011)}.


\bibitem{Unified-SMG}
P. Brax, A.C. Davis, B. Li, and H.A. Winther,
\href{http://dx.doi.org/10.1103/PhysRevD.86.044015}{Phys. Rev. D {\bf 86}, 044015 (2012)}.
\bibitem{SMG}
P. Brax,
\href{http://inspirehep.net/record/1203611}{Acta Phys. Polon. B {\bf43}, 2307 (2012)};
\href{http://dx.doi.org/10.1103/PhysRevD.90.023505}{Phys. Rev. D {\bf90}, 023505 (2014)};
\href{http://dx.doi.org/10.1088/0264-9381/30/21/214005}{Class. Quant. Grav. {\bf30}, 214005 (2013)};
A.C. Davis, R. Gregory, R. Jha, and J. Muir,
\href{http://dx.doi.org/10.1088/1475-7516/2014/08/033}{JCAP {\bf1408}, 033 (2014)}.


\bibitem{donato}
F. Donato, G. Gentile, P. Salucci, C. Frigerio Martins, M. I. Wilkinson, G. Gilmore, E. K. Grebel, A. Koch and R. Wyse,
\href{http://dx.doi.org/10.1111/j.1365-2966.2009.15004.x}{MNRAS {\bf397}, 1169 (2009)}.

\bibitem{SMG-cluster}
M. Gronke, D.F. Mota, and H.A. Winther,
\href{http://dx.doi.org/10.1051/0004-6361/201526611}{Astron. Astrophys. {\bf583}, A123 (2015)};
A. Terukina, L. Lombriser, and K. Yamamoto,
\href{http://dx.doi.org/10.1088/1475-7516/2014/04/013}{JCAP {\bf1404}, 013 (2014)}.


\bibitem{ppn-st-gen-potential}M. Hohmann, L. J\"arv, P. Kuusk, and E. Randla, \href{http://dx.doi.org/10.1103/PhysRevD.88.084054}{Phys. Rev. D {\bf 88}, 084054 (2013)}; \href{http://dx.doi.org/10.1103/PhysRevD.89.069901}{{\bf 89}, 069901(E) (2014)}.


\bibitem{Testing-st-ppn}A. Sch\"arer, R. Angelil, R. Bondarescu, P. Jetzer, and A. Lundgren, \href{http://dx.doi.org/10.1103/PhysRevD.90.123005}{Phys. Rev. D {\bf 90}, 123005 (2014)}.


\bibitem{Will:1993ns}C.M. Will,\textit{Theory and Experiment in Gravitational Physics},(Cambridge University Press, Cambridge, 1993).


\bibitem{WillReview}C.M.~Will, \href{http://www.livingreviews.org/lrr-2014-4} {Liv. Rev. Rel. {\bf 17},  4 (2014)}.


\bibitem{Beyond Cos-Sta-Mod}
A. Joyce, B. Jain, J. Khoury, and M. Troddenb, \href{http://dx.doi.org/10.1016/j.physrep.2014.12.002}{Phys. Rept. {\bf568}, 1 (2015)}.


\bibitem{Combined-cc-ss} A.~Hees and A. F\"uzfa, \href{http://dx.doi.org/10.1103/PhysRevD.85.103005}{Phys.~Rev.~D {\bf 85}, 103005 (2012)}.




\bibitem{nature-gamma}B. Bertotti, L. Iess, and P. Tortora, \href{http://www.nature.com/nature/journal/v425/n6956/full/nature01997.html}{Nature (London) {\bf 425}, 374 (2003)}.


\bibitem{f(R)-S-AdS}E. Babichev and D. Langlois, \href{http://dx.doi.org/10.1103/PhysRevD.81.124051}{Phys. Rev. D {\bf81}, 124051 (2010)}.


\bibitem{2008A&A...477..657A} H.M. Antia, S.M. Chitre, and D.O. Gough,
\href{http://dx.doi.org/10.1051/0004-6361:20078209}{Astron. Astrophys. {\bf 477}, 657 (2008)}.




\bibitem{PIAU} D. Hobbs, B. Holl, L. Lindegren, F. Raison, S. Klioner, and A. Butkevich, \href{http://dx.doi.org/10.1017/S1743921309990561}{Proc. Int. Astron. Union {\bf5}, 315 (2009)}.


\bibitem{milani02} A. Milani, D. Vokrouhlick{\'{y}}, D. Villani, C. Bonanno, and A. Rossi,
\href{http://dx.doi.org/10.1103/PhysRevD.66.082001}{Phys. Rev. D {\bf 66}, 082001 (2002)}.


\bibitem{2007PhRvD..75b2001A} N. Ashby, P.L. Bender,  and J.M. Wahr,
\href{http://dx.doi.org/10.1103/PhysRevD.75.022001}{Phys. Rev. D {\bf 75}, 022001 (2007)}.


\bibitem{Planck 2013 Cosmological parameters} Planck Collaboration, \href{http://dx.doi.org/10.1051/0004-6361/201321591}{Astron. Astrophys. {\bf 571},  A16 (2014)}.


\bibitem{stumwshr2007}C.~Schimd, I.~Tereno, J.P.~Uzan, Y.~Mellier, L.~van~Waerbeke, E. Semboloni, H. Hoekstra, L. Fu, and A.~Riazuelo, \href{http://dx.doi.org/10.1051/0004-6361:20065154}{Astron. Astrophys. {\bf 463}, 405 (2007)}.



\end{thebibliography}
\end{document}